\documentclass[aps,prb,reprint,superscriptaddress,longbibliography]{revtex4-2}
\usepackage{amsmath}
\usepackage{amssymb}
\usepackage{CJKutf8}
\usepackage{graphicx}
\usepackage{dcolumn}
\usepackage{bm}
\usepackage{hyperref}
\usepackage[dvipsnames]{xcolor}
\usepackage[tight]{subfigure} 
\usepackage{mathtools}
\usepackage{bbold}
\usepackage{ulem}
\usepackage{tikz}
\usepackage{esint}
\usepackage{multirow}
\usepackage{float}
\usetikzlibrary{positioning}

\newcommand{\be}{\begin{equation}}
\newcommand{\ee}{\end{equation}}
\newcommand{\C}{\mathcal{C}}
\renewcommand{\L}{\mathcal{L}}
\newcommand{\vphi}{\varphi}

\newcommand{\Z}{\mathbb{Z}}

\newcommand{\p}{\partial}

\begin{document}
\begin{CJK*}{UTF8}{}

\title{Boundary theory of the X-cube model in the continuum}

\author{Zhu-Xi Luo (\CJKfamily{gbsn}罗竹悉)}
\affiliation{Kavli Institute for Theoretical Physics, University of California, Santa Barbara, CA 93106, USA}

\author{Ryan C. Spieler}
\author{Hao-Yu Sun  (\CJKfamily{bsmi}孫昊宇)}
\author{Andreas Karch}
\affiliation{Department of Physics, University of Texas, Austin, TX 78712-1192, USA}

\date{\today}

\begin{abstract}
We study the boundary theory of the $\Z_N$ X-cube model using a continuum perspective, from which the exchange statistics of a subset of bulk excitations can be recovered. We discuss various gapped boundary conditions that either preserve or break the translation/rotation symmetries on the boundary, and further present the corresponding ground state degeneracies on $T^2\times I$. The low-energy physics is highly sensitive to the boundary conditions: even the extensive part of the ground state degeneracy can vary when different sets of boundary conditions are chosen on the two boundaries. We also examine the anomaly inflow of the boundary theory and find that the X-cube model is not the unique (3+1)d theory that cancels the 't Hooft anomaly of the boundary.
\end{abstract}


\maketitle
\end{CJK*}

\tableofcontents

\section{Introduction}
\label{sec:intro}

Fracton phases constitute a new class of quantum matter, which involve emergent quasiparticles with constrained mobility \cite{Chamon,BRAVYI2011839, Haah,Yoshida,PhysRevB.92.235136,Vijay_original, Rahul_review,Pretko_review}: they can either move only in certain directions (type-I), or cannot move at all without creating additional excitations (type-II).  Similar to topologically ordered phases, (3+1)d gapped fracton phases exhibit robust ground state degeneracies (GSDs) when defined on nontrivial manifolds, i.e., manifolds with non-contractible cycles. However, unlike topological orders where the the GSDs are constants, fracton phases often host extensive GSDs which grows exponentially with the linear size of the system.

Realistic samples are naturally finite systems with open boundaries. Experiences from conventional topological orders show that essential  topological properties of the bulk are encoded in their boundary theories, i.e., there is a bulk-boundary correspondence. Examples include the fractional statistics of emergent bulk quasiparticles, transport properties and entanglement entropy, etc. \cite{PhysRevB.25.2185,cmp/1104178138,wen1992theory,PhysRevLett.71.3697,CAPPELLI2002568,CAPPELLI2010465,Cappelli_2011,PhysRevB.94.045113}.  
Many fracton phases of matter can be constructed by stacking and coupling conventional (2+1)d topological orders \cite{MaHan,PhysRevX.8.031051,KevinFoliated,10.21468/SciPostPhys.6.4.041,SHIRLEY2019167922,PhysRevLett.126.101603}. Most recently, it has also been conjectured that all gapped fracton phases of matter can be constructed using networks of defects in (3+1)d topological orders \cite{PhysRevResearch.2.043165, https://doi.org/10.48550/arxiv.2112.14717}. It is thus natural to ask whether there exists a bulk-boundary correspondence in fracton phases of matter, similar to that in topological phases of matter. 

The boundary theories for fracton phases might be more exotic and interesting because the systems are very sensitive to boundary geometries. However, the studies of the fracton boundaries are sparse. In Refs. \cite{Daniel, Screw}, the authors focused on the $\mathbb{Z}_2$ X-cube model \cite{Vijay_original}, which is the canonical  example for type-I or foliated fracton orders \cite{SHIRLEY2019167922,PhysRevB.99.115123}. Both works analyzed certain typical gapped boundary conditions on the lattice: \cite{Daniel} focused on the boundary excitations and mobility constraints, while \cite{Screw} is concerned with possible dislocations in the bulk. 

In this work, we instead take the perspective of continuum theories  \cite{Cenke,gu2006lattice,PhysRevD.81.104033,GU201290,rasmussen2016stable,Pretko1, Pretko2,KevinKim,bulmash2018generalized,PhysRevX.9.031035, PhysRevB.101.085106,Yau,Seiberg,YOU2020168140,SS1,SS2,SS3,SS4,PhysRevResearch.2.023249,fontana2021field}, which typically render the symmetries and universal properties of the system more manifest. We will focus on the $\mathbb{Z}_N$ version of the X-cube model \cite{PhysRevB.96.195139,KevinKim,SS3,Jintae} and its boundaries perpendicular to the $(001)$-direction. The boundary theory encodes the braiding statistics of certain bulk excitations and is governed by a generalization of the $U(1)$ Kac-Moody algebra. Various gapping terms are considered, which not only match with all gapped boundaries found in ref. \cite{Daniel} from the lattice perspective, but also include additional gapped boundary conditions not present there. We then count the GSDs with these boundary conditions on the $T^2\times I$ geometry. Interestingly, even the extensive part in the GSD changes with different boundary conditions. 

To further understand the relation between the boundary and bulk theories, we examine the anomaly inflow \cite{CALLAN1985427}. The boundary theory itself is anomalous in the sense that it cannot be consistently coupled to background gauge fields. We ask whether this anomaly uniquely determines the bulk theory to be the X-cube model, and the answer is negative. We present a simple $(3+1)$d theory that cancels the boundary anomaly and is distinct from the X-cube model. This result is closely related to the fact that certain exchange statistics that involves the movement of the bulk quasiparticles in the third spatial direction perpendicular to the boundary cannot be recovered purely from that boundary. The non-uniqueness of bulk theories given a boundary theory is expected to be common in fracton phases of matter \cite{anomaly,private}. 

The remainder of this work is organized as follows. We first review the bulk continuum theory of the $\mathbb{Z}_N$ X-cube model in Section \ref{sec:review}. Then we derive its boundary Lagrangian, discuss its symmetries, spectrum and indications of exchange statistics of the bulk excitations. Later in sections  \ref{sec:mm_mm}, \ref{sec:ee_ee}, \ref{sec:ee_mm} and \ref{sec:em_em}, we examine several simple examples of gapped boundary conditions and count their GSDs from both the continuum, and present alternative lattice countings as consistency checks. Next we study more general gapped boundary conditions in Section \ref{sec:Z_N}, including those that arise from mixtures of elementary boundary conditions. In Section \ref{sec:anomaly}, we examine the anomaly inflow, and then end with a discussion in Section \ref{sec:discussion}.

\section{Review of the continuum description of the X-cube model on \texorpdfstring{$T^3$}{}}
\label{sec:review}
We begin by recapitulating the effective field theory for the X-cube model \cite{KevinKim,SS3} and will largely follow the conventions in \cite{SS3}. We will mostly work in Euclidean signature unless otherwise stated. 

Since the X-cube model is defined on a three-dimensional cubic lattice, we will focus on the orientation-preserving subgroup of the cubic group, which is $S_4$. In particular, we are interested in the gauge fields $(A_0, A_{ij})$ which furnish the $(\textbf{1},\textbf{3}')$ irreducible representations of $S_4$,  and the gauge fields $(\hat{A}_0^{k(ij)},\hat{A}^{ij})$ in the $(\textbf{2},\textbf{3}')$ irreducible representations of $S_4$. (See Appendix \ref{app:repn} for details of the representations.) They have the following gauge transformations: 
\begin{equation}
\begin{split}
    A_0 &\sim A_0 + \partial_0 \alpha,\quad\quad\quad\quad A_{ij} \sim A_{ij} + \partial_i \partial_j \alpha,\\
     \hat{A}_0^{i(jk)} &\sim \hat{A}_0^{i(jk)} + \partial_0 \hat{\alpha}^{i(jk)},\quad \hat{A}^{ij} \sim \hat{A}^{ij} + \partial_k \hat{\alpha}^{k(ij)},
\end{split}
\label{eq:gauge_bulk}
\end{equation}
where repeated indices are summed over, $\alpha$ is a $2\pi$-periodic scalar, $\hat{\alpha}^{k(ij)}$  lies in the representation $\textbf{2}$ of $S_4$ and is also $2\pi$-periodic.  These gauge transformations preserve the following field strengths: 
\begin{equation}
\begin{split}
   & E_{ij} = \partial_0 A_{ij} - \partial_i \partial_j A_0,\quad 
   B_{k(ij)}=\partial_{[k}A_{i]j}+\partial_{[k}A_{j]i},\\
   & \hat{E}^{ij} = \partial_0 \hat{A}^{ij} - \partial_k \hat{A}_0^{k(ij)},\quad \hat{B} = \frac{1}{2}\partial_i \partial_j \hat{A}^{ij}.
\end{split}
\label{eq:strength_bulk}
\end{equation}
Again, repeated indices are summed over.

In Euclidean signature, the X-cube model is a BF theory: a product of the gauge fields $(A_0,A_{ij})$ with the field strengths for $(\hat{A}_0^{k(ij)},\hat{A}_{ij})$, or vice versa:
\begin{equation}
    \mathcal{L} = i\frac{N}{4\pi} \big[A_0\big(\partial_i\partial_j \hat{A}^{ij}\big)+A_{ij}\big(\partial_0 \hat{A}^{ij} -\partial_k \hat{A}_0^{k(ij)}\big) \big].
    \label{eq:BF1}
\end{equation}
The equations of motion enforce vanishing of all four field strengths in \eqref{eq:strength_bulk}, ensuring that no nontrivial local gauge-invariant operators exist. However, the theory contains non-local gauge-invariant operators that are analogues of Wilson lines.  They can be constructed from either  $(A_0, A_{ij})$ or  $(\hat{A}_0^{k(ij)},\hat{A}^{ij})$ fields.  These operators can be viewed as probe limits of the massive excitations of the X-cube; we simply refer to them as ``excitations'' hereafter.

The simplest defect constructed from the $(A_0, A_{ij})$ fields is a charge at a spatial point
\begin{equation}
    W = \exp \left[i\int_{-\infty}^\infty dt\ A_0(t,x,y,z)\right]
\end{equation}
which describes a single, immobile fracton with gauge charge $+1$ at a fixed point in space. Certain composites of such fractons are, however, mobile. This is encapsulated by the operators of the form
\begin{equation}
\begin{split}
    & W_{xy}(z_1,z_2,\mathcal{C}_{xy}) \\
    = & \exp \left[i\int_{z_1}^{z_2}dz\int_{\mathcal{C}_{xy}} (dt \partial_z A_0 + dx A_{xz} +dy A_{yz}) \right]
\end{split}
\label{eq:planon}
\end{equation}
which describe a dipole of fractons with gauge charges $\pm 1$, separated in the $z$-direction, and moving along a curve $\mathcal{C}_{xy}$ in the $(t,x,y)$ hyperplane. Since such a dipole can freely move in the spatial $xy$-plane, it is called a ``planon''. (We omit $t$ in the subscript of $\C_{xy}$ to emphasize the spatial mobility of this dipole.)   
Similarly, we can construct gauge-invariant operators in the other spatial planes as well. 

Turning to the $(\hat{A}_0^{k(ij)},\hat{A}^{ij})$ fields, the simplest defect with gauge charge $+1$ one can write down is
\begin{equation}
    \hat{W} = \exp \left[i\int_{-\infty}^\infty dt\ \hat{A}_0^{i(jk)}\right],   
\end{equation}
which is mobile in one spatial direction. For example one can write the following defect operator 
\begin{equation}
    \hat{W}^z(x,y,\hat{\mathcal{C}}_z) = \exp\left[i\int_{\hat{\mathcal{C}}_z} (dt \hat{A}^{z(xy)}_0 + dz\hat{A}^{xy})\right],
\label{eq:lineon}
\end{equation}
where $\hat{\mathcal{C}}_z$ is a spacetime curve in the $(t,z)$ plane. The expression above describes a ``lineon'' that can only move along a spatial line in the $z$-direction, which we call a ``$z$-lineon''. A dipole of lineons can be further combined to form planons, such as: 
\begin{equation}
\begin{split}
     \hat{W}^{zy}(x_1,x_2,\hat{\C}_{yz}) 
    &= \exp \left[i\int_{x_1}^{x_2}dx \int_{\hat{\C}_{yz}}   \left( dt\ \partial_x \hat{A}^{z(xy)}_0\right.\right. \\
    & \,\,\left.\left.+\ dz\ \partial_x \hat{A}^{xy}-dy\ (\partial_x\hat{A}^{xz} +\partial_y \hat{A}^{yz}) \right) \right],
\end{split}
\label{eq:e_planon}
\end{equation}
where a dipole of $z$-lineons \eqref{eq:lineon} are separated in the $x$-direction, located at $x_1$ and $x_2$, respectively, and $\hat{\C}_{yz}$ is a spacetime curve in the $(t,y,z)$-plane. 

We will call the defects described by $\hat{W}$'s the electric ones, and those described by $W$'s the magnetic ones, as they exhibit mixed anomalies which will be discussed soon. Their names should not be confused with the $(E,B)$ and $(\hat{E},\hat{B})$ fields in \eqref{eq:strength_bulk}. 
The restricted mobilities of these defects are reflected in the fact that $\hat{\mathcal{C}}$ and $\mathcal{C}$  are constrained in subdimensional space, indicative of the foliation structure of the theory and in stark contrast to the analogous story in topological field theories. One useful observation from equations \eqref{eq:planon} and \eqref{eq:e_planon} is that, magnetic/electric planons can be formed by a dipole of fractons/lineons separated in the $x_i$-direction, and can only move on the plane perpendicular to $x_i$. 

In the remainder of this section we quickly review the counting of ground state degeneracy (GSD) when the field theory is regularized on a lattice on $T^3$. Part of this material will be repeated again when encountered in latter sections.

For any point $(x,y)$ in the $xy$-plane at fixed time, one can define the $\mathbb{Z}_N$ tensor symmetry operator,
\begin{equation}
     \hat{W}^z(x,y) = \exp\left[i\oint dz \hat{A}^{xy}(x,y,z)\right],
\label{eq:tensor_bulk}
\end{equation}
which is a special case of \eqref{eq:lineon}  where $\hat{\mathcal{C}}_z$ forms a closed loop around the $z$-direction. 
Using the fact that $\hat{B}=0$, one can derive $\partial_x\partial_y\oint dz \hat{A}^{xy}=0$, such that the spatial dependence of $\hat{W}^z(x_0,y_0)$ factorizes:
\begin{equation}
\hat{W}^z(x,y)=\hat{W}^z_x(x)\hat{W}^z_y(y).
\label{eq:tensor_split_bulk}
\end{equation}
Thus, there is a $U(1)$ gauge redundancy between the two: multiplying $\hat{W}^z_x$ by a phase factor and $\hat{W}^z_y$ by the conjugate phase, $\hat{W}^z(x,y)$ is invariant.

In addition, one can define the $\mathbb{Z}_N$ dipole winding operators 
{\small \begin{equation}
\begin{split}
& W_{yz,y}(x_1,x_2,\mathcal{C}_{yz,y}) = \exp\left[i\int_{x_1}^{x_2} dx\oint_{\mathcal{C}_{yz,y}}dy A_{xy} + dz  A_{xz}\right]\\
& W_{xz,x}(y_1,y_2,\mathcal{C}_{xz,x}) = \exp\left[i\int_{y_1}^{y_2} dy\oint_{\mathcal{C}_{xz,x}}dx A_{xy}+dz A_{yz}\right],
\end{split}
\label{eq:dipole_wind_bulk}
\end{equation}}
\hspace{-3.4pt}which are the special cases of \eqref{eq:planon}, where $\mathcal{C}_{ij,i}$ is a curve in the $ij$ plane that wraps once around the non-contractible $i$ direction but not along the $j$ direction. 

The operators defined above \eqref{eq:tensor_bulk} \eqref{eq:dipole_wind_bulk} obey the following commutation relations,
\begin{equation}
\begin{split}
    & \hat{W}^z(x,y) W_{yz,y}(x_1,x_2,\mathcal{C}_{yz,y})  \\
    =\ & e^{2\pi i/N}  W_{yz,y}(x_1,x_2,\mathcal{C}_{yz,y})\hat{W}^z(x,y),\\
    & \hat{W}^z(x,y) W_{xz,x}(y_1,y_2,\mathcal{C}_{xz,x})\\
    =\ & e^{2\pi i/N} W_{xz,x}(y_1,y_2,\mathcal{C}_{xz,x})\hat{W}^z(x,y),
\end{split}
\label{eq:commut_bulk}
\end{equation}
with $x\in (x_1,x_2), y\in (y_1,y_2)$. When regularized on a lattice with $l_x\times l_y\times l_z$ sites, these relations \eqref{eq:commut_bulk} form an algebra isomorphic to $l_x+l_y-1$ copies of the $\mathbb{Z}_N$ Heisenberg algebra, with the $-1$ coming from the constraint 
\be W_{yz,y}(0,L_x,\C_{yz,y})= W_{xz,x}(0,L_y,\C_{xz,x}).
\label{eq:W_constraint}
\ee
$L_i=al_i$ with $a$ being the UV cutoff (lattice constant), and $L_i$ the linear size of the system in the $x_i$ direction. Equation \eqref{eq:commut_bulk} makes manifest the mixed anomaly between the two $\Z_N$ subsystem symmetries generated by the $\hat{W}^z$ and $W_{xz}$, $W_{yz}$ operators when they intersect respectively. Accounting for similar relations in other directions as well leads to
\begin{equation}
    GSD_{T^3} = N^{2l_x+2l_y+2l_z-3}.
\end{equation}
For later convenience, we summarize the counting of ground state degeneracy in the table below.
\setlength{\tabcolsep}{5pt}
\renewcommand{\arraystretch}{1.5}
\begin{table}[htbp]
    \centering
    \begin{tabular}{|c|c|c|}
    \hline
       \vtop{\hbox{\strut Non-commuting}\hbox{\strut \quad operators}}  &  \vtop{\hbox{\strut \quad Copies of}\hbox{\strut Heisenberg alg.}}   & \vtop{\hbox{\strut Contribution}\hbox{\strut \quad to GSD}}\\ \hline
       $\hat{W}^x$, $W_{xz,z}$  &  $l_y$ & \multirow{2}{*}{$N^{l_y+l_z-1}$} \\ \cline{1-2}
       $\hat{W}^x$, $W_{xy,y}$  &  $l_z$ & \\ \hline
       $\hat{W}^y$, $W_{yz,z}$  &  $l_x$ & \multirow{2}{*}{$N^{l_x+l_z-1}$} \\ \cline{1-2}
       $\hat{W}^y$, $W_{xy,x}$  &  $l_z$ & \\ \hline
       $\hat{W}^z$, $W_{yz,y}$  &  $l_x$ & \multirow{2}{*}{$N^{l_x+l_y-1}$} \\ \cline{1-2}
       $\hat{W}^z$, $W_{xz,x}$  &  $l_y$ & \\ \hline
    \end{tabular}
    \caption{Summary of Wilson operators and their contributions to the ground state degeneracy.}
    \label{tab:T3}
\end{table}
One can also arrive at the same ground state degeneracy by considering the nontrivial commutation relations between the planon operators in the electric sector $\hat{W}^{ij}$ and those in the magnetic sector $W_{ij}$. We will, however, not take this viewpoint as $\hat{W}^{ij}$ are not the elementary mobile electric excitations.

\section{Adding boundaries}
\label{sec:bdry}

The BF theory \eqref{eq:BF1} for the X-cube model is not gauge invariant on manifolds with boundary, similar to the case of (2+1)d Chern-Simons theory.  To see this, consider the X-cube model with a toroidal boundary at $z=0$.  A gauge transformation $\alpha$ changes the action in the following way:
\begin{equation}
\begin{split}
    S \rightarrow S-  \frac{iN}{2\pi}\int_{z=0} & d\tau dx dy\  [(\partial_x \partial_y \alpha) \hat{A}^{z(xy)} \\
    & + (\partial_0 \partial_y \alpha) \hat{A}^{yz} + (\partial_0 \partial_x \alpha) \hat{A}^{xz} ].
\end{split}
\end{equation} 
To retain gauge invariance, we follow the strategy used in Chern-Simons theories (see for example \cite{MOORE1989422, WenBoundary,Tong,Subir}) and only allow the gauge transformations that vanish on the boundary.  By restricting gauge transformations, we are introducing new degrees of freedom that live on the boundary.  In this section, we derive and analyze the action for those degrees of freedom.  

To start, note that the the boundary at $z=0$ adds yet another complication.  The variation of the action has another term along the boundary:   
\be
    \delta S\hspace{-3pt}\mid = -\frac{iN}{2\pi}\int_{\p \mathcal{M}}d^3x [\delta A_{xy} \hat{A}_0^{z(xy)} + \delta A_0 (\partial_y  \hat{A}^{yz} + \partial_x \hat{A}^{xz})].
\ee
To preserve the bulk equations of motion, we force $\delta S\hspace{-3.5pt}\mid$ to vanish by imposing the boundary conditions: 
\be
A_0\mid=0,\quad \hat{A}_0^{k(ij)}\mid=0.
\label{eq:temporal}
\ee
We can extend this boundary condition into the bulk as a gauge choice, i.e., by removing the ``\ $\mid$\ '' in the equations above, leading to the following constraints 
\be
\hat{B}=0,\quad B_{k(ij)}=0. 
\ee
Recalling \eqref{eq:strength_bulk}, solutions to these equations of motion  are simply
\be
A_{ij}=\p_i \p_j \vphi,\quad
\hat{A}^{ij}=\p_k \hat{\vphi}^{k(ij)},
\label{eq:A_phi}
\ee
where $\hat{\vphi}^{k(ij)}$ is in the irreducible representation $\bm{2}$ of $S_4$, satisfying $i, j, k,$ not equal to each other and $\hat{\vphi}^{x(yz)}+\hat{\vphi}^{y(zx)}+\hat{\vphi}^{z(xy)}=0.$  

In the following, we will simplify the notation $\hat{\varphi}^{k(ij)}\equiv \hat{\varphi}^{k}.$ 
Inserting the above solutions back into the bulk action, we find that the only nonzero contributions come from the boundary,
\be
\L_0=-i\frac{N}{2\pi} \left[(\partial_x \partial_y \vphi)\ \partial_0 \hat{\vphi}^{y}+(\partial_x \partial_y \vphi)\ \partial_0 \hat{\vphi}^{x}\right].
\label{eq:boundary}
\ee
Since the $z=0$ boundary is a square lattice in the UV, the orientation-preserving subgroup of the boundary symmetry group is $\mathbb{Z}_4$. $\varphi$ transforms in the trivial representation of $\mathbb{Z}_4$.  $\hat{\varphi}^{z}=-\hat{\varphi}^{x}-\hat{\varphi}^{y}$ transforms 
in the $\bm{1}_2$ representation, while $\hat{\varphi}^{x}-\hat{\varphi}^{y}$ transforms in the $\bm{1}_0$ representation of $\Z_4$. See Appendix \ref{app:repn} for details. 

These fields are compact, subject to the following identifications on the boundary: 
\begin{equation}
\begin{split}
   \vphi(\tau,x,y) &\sim \vphi(\tau,x,y) + 2\pi w_x(x) + 2\pi w_y(y),\\
    \hat{\vphi}^x(\tau,x,y) &\sim \hat{\vphi}^x(\tau,x,y) -2\pi \hat{w}^y(y),\\
    \hat{\vphi}^y(\tau,x,y) &\sim \hat{\vphi}^y(\tau,x,y) + 2\pi \hat{w}^x(x),\\
  \hat{\vphi}^z(\tau,x,y) &\sim \hat{\vphi}^z(\tau,x,y) -2\pi \hat{w}^x(x) + 2\pi \hat{w}^y(y).
\end{split}
\end{equation}
where we have used the identity $\hat{\vphi}^z=-\hat{\vphi}^x-\hat{\vphi}^y$. $w_x$, $w_y$, $\hat{w}^x$, and $\hat{w}^y$ are all integer-valued functions.  Operators such as $\partial_x\partial_y \vphi$, $e^{i\vphi}$,  $\partial_x \hat{\vphi}^x$, $\partial_y \hat{\vphi}^y$, $\partial_x \partial_y \hat{\vphi}^z$, and $e^{i\hat{\vphi}^i}$ survive the identification whereas operators such as $\vphi$, $\partial_x \vphi$, $\partial_y \vphi$, $\hat{\vphi}^i$, $\partial_x \hat{\vphi}^z$, and $\partial_y \hat{\vphi}^z$ do not. 

Notice the fact that only $\hat{\vphi}^z$ appears in $\L_0$ means that the dynamics is governed by $\hat{\vphi}^{x}+\hat{\vphi}^{y}$, not the other combination of $\hat{\vphi}^{x}-\hat{\vphi}^{y}$. But the degree of freedom $\hat{\vphi}^x-\hat{\vphi}^y$ is still present in the system and can have nontrivial winding configurations. 

Using the identity $\hat{\vphi}^z=-\hat{\vphi}^x-\hat{\vphi}^y$ and defining for convenience $\bm{\Phi}=(\Phi_1,\Phi_2)=(\vphi,\hat{\vphi}^z)$, the Lagrangian can be written in a more compact form 
\be
\L_0=\frac{i K_{IJ}}{4\pi}\p_0\Phi_I \p_x\p_y\Phi_J,\quad K=-iN\sigma^y,\quad I,J=1,2,
\ee
where the summation over $I, J$ is implicit. Notice that $K$ is anti-symmetric because of the double spatial derivatives above. This boundary theory exhibits the following momentum subsystem symmetries
\be
\Phi_I(t,x,y)\rightarrow \Phi_I(t,x,y)+f_I(x,y),
\label{eq:enhanced}
\ee
with $f_I$ an arbitrary function of $x$ and $y$. These symmetries are generated by the currents
\be
J_{I,0}=-\frac{K_{IJ}}{2\pi}\p_x\p_y\Phi_J,\quad J_{I,xy}=0.
\ee
From the equation of motion, $\p_0 J_{I,0}=0$ and consequently the charge is conserved everywhere on the surface. The symmetry \eqref{eq:enhanced} is in general broken by local gauge-invariant terms, such as $A_{xy}^2$, $(\p_x \hat{A}^{xz}+\p_y \hat{A}^{yz})^2$ and $A_{xy}(\p_x \hat{A}^{xz}+\p_y \hat{A}^{yz})$, that can be added on the boundary.
We thus arrive at the following general Lagrangian, presented in the Lorentizan signature, 
\be
\L_{\p \mathcal{M}} = \frac{1}{4\pi}\left[K_{IJ} \p_0\Phi_I \p_x\p_y \Phi_J-V_{IJ}(\p_x\p_y\Phi_I)  (\p_x\p_y\Phi_J)\right],
\label{eq:boundary_full}
\ee
where $V_{IJ}$ is a non-universal, constant matrix, which is positive-definite as it is the only term in the Hamiltonian. $V$ thus resembles the velocity matrix of the boundary theory of the (2+1)d $K$-matrix Chern-Simons theory \cite{WenBoundary,Juven,Levin,ChaoMing}. The remaining symmetries are momentum dipole symmetries,
\be
\Phi_I\rightarrow \Phi_I+f_{I,x}(x)+f_{I,y}(y)
\label{eq:momentum_dipole}
\ee
with $f_{I,i}(x_i)$ being arbitrary functions of $x_i$. The corresponding currents are 
\begin{equation}
\begin{split}
 J_I^0 &= \frac{K_{IJ}}{4\pi}\partial_x \partial_y \Phi_J,\\
 J_I^{xy} &= -\frac{K_{IJ}}{4\pi}\p_0\Phi_J-\frac{(V_{IJ}+V_{JI})}{4\pi}\partial_x \partial_y \Phi_J,
\end{split}
\end{equation}
satisfying $\p_0 J_I^0=\p_x\p_y J_I^{xy}$. 
Plugging in the plane-wave ansatz $\Phi_I=C_I e^{i\omega t+i\vec{k}\cdot \vec{x}}$ into the equation of motion with $C_I$ being a constant, we arrive at 
the following dispersion relation 
\be
\omega^2=[4V_{11}V_{22}-(V_{12}+V_{21})^2]k_x^2k_y^2/4N^2,
\label{eq:dispersion_gapless}
\ee 
which is gapless.  The system also exhibits the following dipole winding symmetries,
\be
\tilde{J}_{I,0}=\frac{N}{2\pi}\p_x\p_y\Phi_I,\quad \tilde{J}_{I,xy}=\frac{N}{2\pi}\p_0 \Phi_I,
\ee
satisfying $\p_0 \tilde{J}_{I,0}=\p_x\p_y \tilde{J}_{I,xy}$. The winding symmetries of $\hat{\vphi}^x$ and $\hat{\vphi}^y$ are inherited from the bulk theory. 

To examine the boundary excitations and statistics, we identify the charge densities on the boundary as (The coupling of the boundary theory to background tensor gauge fields are presented in Appendix \ref{sec:background} in the spirit of ref. \cite{anomaly}.),
\be
\rho_I=\frac{1}{2\pi}\p_x\p_y\Phi_I. 
\ee
They constitute the boundary Hamiltonian and govern the edge dynamics, 
\be
\mathcal{H}_{\p \mathcal{M}}=\pi V_{IJ}\rho_I\rho_J. 
\ee
Next, we define the vertex operators
\be
V_I=e^{i\Phi_I},\quad V_I^\dagger=e^{-i\Phi_I},
\label{eq:V}
\ee
which satisfy the following commutation relations with the charge densities
\be 
\begin{split}
& [\rho_I(\vec{x}),V_J^\dagger(\vec{x}')]=(K^{-1})_{IJ} \delta^{(2)}(\vec{x}-\vec{x}') V_I^\dagger(\vec{x}')\\
& [\rho_I(\vec{x}),V_J(\vec{x}')]=-(K^{-1})_{IJ} \delta^{(2)}(\vec{x}-\vec{x}') V_I(\vec{x}').
\end{split}
\label{eq:vertex}
\ee 
Therefore, $V_I^\dagger$ or $V_I$ should be interpreted as the creation and annihilation operators of the fractons or $z$-lineons (corresponding to the violations of $\hat{B}=0$ or $B_{z(xy)}=0$, respectively). Equation \eqref{eq:vertex} can be seen by performing the Fourier expansion in real space, 
\be
\Phi_I(\vec{x})=
(L_x L_y)^{-1/2}
\sum_{\vec{k}} e^{i\vec{k}\cdot \vec{x}}\  \Phi_{I,\vec{k}},
\ee 
where $k_i=2\pi n_i/l_i$ with $n_i\in \Z$. One can then read off from the Lagrangian
\be
[\Phi_{I,\vec{k}}, \Phi_{J,\vec{k}'}]=2\pi i \frac{(K^{-1})_{IJ}}{k_x k_y}\delta_{\vec{k},-\vec{k}'},
\label{eq:commutation_k}
\ee
where the $\delta$ above is the Kronecker delta. Multiply both sides by $k_x k_y$ and transform back into the real space,
\be
[\rho_I(\vec{x}),\Phi_J(\vec{x}')]= i(K^{-1})_{JI}\delta^{(2)}(\vec{x}-\vec{x}').
\ee
One then arrives at \eqref{eq:vertex} upon exponentiation. It can also be easily seen that $\rho_{I\vec{k}}$'s form an analog of the $U(1)$ Kac-Moody algebra, 
\be
[\rho_{I\vec{k}},\rho_{J\vec{k}'}]= \frac{i (K^{-1})_{IJ}}{2\pi}k_x k_y \delta_{\vec{k}+\vec{k}'}.
\ee

The mutual statistics of these boundary excitations can be further computed:
\be 
V_I(\vec{x})V_J(\vec{x}')=e^{-[\Phi_I(\vec{x}),\Phi_J(\vec{x}')]} V_J(\vec{x}')V_I(\vec{x}).
\ee
The commutator can be again obtained from \eqref{eq:commutation_k}:
\be
[\Phi_{I}(\vec{x}),\Phi_J(\vec{x}')]
=\frac{i\pi}{2} (K^{-1})_{JI} \text{sgn}(x-x')\ \text{sgn}(y-y').
\label{eq:weird}
\ee
However, the above equation \eqref{eq:weird} should not be understood as the phase arising from the exchange of a fracton and a lineon, because of the restricted mobility of the two: a single fracton is immobile and therefore cannot wind around a $z$-lineon, while a $z$-lineon can only move along $z$-direction and therefore immobile on the $xy$-boundary. 
One can, however, consider the mutual statistics of a dipole of fractons, which form a magnetic planon, and a single $z$-lineon. For example, taking a dipole of fractons at $(x_1,y_1)$ and $(x_1',y_1')$, and a $z$-lineon at $(x_2,y_2)$, with $x_1<x_2<x_1'$ and $y_1=y_1'$, then
\be
\begin{split}
& V_1(x_1,y_1)V_1^\dagger(x_1',y_1')V_2(x_2,y_2)\\
 = &\ e^{i\theta}  V_2(x_2,y_2)V_1(x_1,y_1)V_1^\dagger(x_1',y_1'),
\end{split}
\label{eq:statistics1}
\ee
where the statistical angle $\theta$ is
\be
\begin{split}
\theta & =[\Phi_1(x_1',y_1'),\Phi_2(x_2,y_2)]-[\Phi_1(x_1,y_1),\Phi_2(x_2,y_2)]\\
& = -\frac{\pi i}{N}\  \text{sgn}(y_1-y_2).
\label{eq:statistics2}
\end{split}
\ee
Alternatively, one can also study the mutual statistics between a dipole of $z$-lineons separated in the $x$ (or $y$-) direction, which is mobile along the $y$ (or $x$-) direction on the $xy$-boundary, and a single fracton. 
For instance, taking a dipole of two $z$-lineons at $(x_2,y_2)$ and $(x_2',y_2')$ and a fracton at $(x_1,y_1)$, with $x_2<x_1<x_2'$ and $y_2=y_2'$, then
\be
\begin{split}
& V_1(x_1,y_1)V_2(x_2,y_2)V_2^\dagger(x_2',y_2')\\
 = &\ e^{-i\theta}  V_2(x_2,y_2)V_2^\dagger(x_2',y_2') V_1(x_1,y_1). 
\end{split}
\ee 

Below we will be interested in the possible gapped boundaries arising from local interactions. The discussions will include nontrivial generalizations of the gapped boundaries in non-chiral gapped topological phases in (2+1)d \cite{Juven,Levin,ChaoMing,PhysRevLett.89.181601,bais2003hopf, BAIS2007552,PhysRevB.79.045316, kitaev2012models,KONG2014436,Hung_Wan,hung2015generalized, Bernevig1, Bernevig2,  hu2018boundary,PhysRevLett.114.076402,Fiona}. For an arbitrary $N$, there always exist at least four gapped boundary conditions, among which two preserve both translation and rotation symmetries of the boundary (the smooth/rough boundaries), and the other two break the fourfold rotation symmetries down to twofold. We will discuss each of them separately in the forthcoming sections, and then present further results regarding more general gapped boundary conditions. For some of the cases where lattice results are also available \cite{Daniel}, our results are consistent with theirs.

\section{Smooth \texorpdfstring{\MakeLowercase{(mm)}}{}  boundaries}
\label{sec:mm_mm}
One simple gapping term is 
\be
\L_g^{(mm)}=g \cos (N\vphi),
\label{eq:mm}
\ee
where the constant $g$ is taken to be very large, ``enforcing'' $\vphi=2\pi m/N$, with $m(x,y)$ being an integer-valued function. 
We are only interested in the possible local interactions that can gap out the boundary, so the perturbative relevance/irrelevance of $\L_g$ is not of interest to us, in the same spirit as the studies of gapped boundary conditions of (2+1)d topological phases \cite{Juven,Levin,ChaoMing}.  
The superscript $(mm)$ is used in \eqref{eq:mm}  because magnetic excitations mobile in either $x$- or $y$-directions in the bulk become condensed on this boundary, i.e., the corresponding Wilson operators are trivial, which will be discussed in more detail later in this section. With \eqref{eq:mm}, the spectrum \eqref{eq:dispersion_gapless} obtains a gap proportional to $\sqrt{g},$ 
\be
\omega^2 = 2g\pi V_{22}+[{4V_{11}V_{22}-(V_{12}+V_{21})^2}]k_x^2k_y^2/4N^2.
\ee

We now discuss the ground state degeneracy on the topology of $T^2\times I$, with two toroidal boundaries perpendicular to the $z$-direction.  $\L_g^{(mm)}$ is added on both boundaries. To obtain a finite ground state degeneracy, we will impose an UV cutoff $a$. For a generic boundary termination, there will be dangling links or open tails with only one endpoint, living on the boundary. So for the $T^2\times I$ topology, with $l_x$, $l_y$ and $l_z$ sites in each direction, the number of links, excluding the open tails, would be $l_x$, $l_y$ and $l_z-1$. Recall $L_i=a l_i$ is the linear size of the system, while $l_i$ is a dimensionless number.  

When boundaries are absent, the gauge-invariant Wilson operators in the system are those presented in table \ref{tab:T3}. In the case with two $(mm)$-type boundaries perpendicular to the $z$-direction, the remaining Wilson operators are shown in table \ref{tab:smooth}. The $\hat{W}^x$ and $\hat{W}^y$ operators are unaffected by the boundaries, while $\hat{W}^z=\exp(i \hat{\vphi}^z\mid^{z_t}_{z_b})$, using the identifications \eqref{eq:A_phi}. $\hat{W}^z$ thus amounts to the creation/annihilation of finite-energy excitations on the boundaries as discussed near \eqref{eq:vertex}, which brings the system out of the ground state subspace and should therefore not be counted. 

For the unhatted, magnetic Wilson operators, we have \be
 W_{xz,z}(y_1,y_2, \mathcal{C}_{xz,z}|)
= \exp\left[i\int_{y_1}^{y_2}dy \int_{z_b}^{z_t} dz\ \p_y \p_z \vphi\right],
\label{eq:m_condense}
\ee
where $\C_{xz,z}\mid$ labels a path that lives in the $xz$-plane and spans the entire $z$-direction of the system, i.e., starts from the bottom boundary $z=z_b$ and ends at the top boundary $z=z_t$; but the path does not wind around the $x$-direction. We have again used the identifications \eqref{eq:A_phi}. Since $\vphi\mid =2\pi m/N$, the expectation values of the $W_{xz,z}(y_1,y_2, \mathcal{C}_{xz,z}|)$ operators are thus $\Z_N$ numbers that depend on $(y_1,y_2)$, forming $l_y$ copies of $\Z_N$ Heisenberg algebras with $\hat{W}^x$. 
Similarly, $W_{yz,z}(x_1,x_2)$ form $l_x$ copies of $\Z_N$ Heisenberg algebras with $\hat{W}^y.$

As for a dipole of fractons separated in the $z$-direction and wind around the $y$-direction in the $xy$-plane, we have \be 
W_{xy,y}(z_1,z_2,\mathcal{C}_{xy,y})
= \exp\left[i\int_{z_1}^{z_2}dz \oint_{\mathcal{C}_{xy,y}} dy\ \p_y \p_z \vphi \right].
\ee
When $z_1\leq z \leq z_2$, such operator form $\mathbb{Z}_N$ Heisenberg algebras with $\hat{W}^x(y,z)$, and there are $l_z$ number of such combinations. However, when $(z_1,z_2)=(z_b,z_t)$, $W_{xy,y}(z_b,z_t,\C_{xy,y})$ is just $W_{xz,z}(0,L_y,\C_{xz,z}|)$ which has already been counted, so 
altogether we have $(l_z-1)$ additional copies of Heisenberg algebra from this set of Wilson operators. The analyses for $W_{xy,x}$ operators can be carried out in parallel.
 
Finally, the $W_{xz,x}$ operator is,  
\be
 W_{xz,x}(y_1,y_2,\mathcal{C}_{xz,x})
=  \exp\left[i\int_{y_1}^{y_2} dy\int_{\mathcal{C}_{xz,x}} dx\ \p_x\p_y \vphi\right]. 
\ee
Such operators are independent of $z$ and evaluates to a finite values on the boundary. This independence of $z$ was discussed in \cite{SS3} and can be directly extended to the case with boundaries,
\be
\begin{aligned}
&\frac{W_{xz,x}(y_1,y_2,\mathcal{C}_{xz,x})}{W_{xz,x}(y_1,y_2,\mathcal{C}_{xz,x}')}\\
&\hspace{60pt}=\exp\left[i\int_{y_1}^{y_2} dy\int_{\mathcal{D}} (\p_x A_{yz}-\p_z A_{xy})\right],
\label{eq:quasi-top}
\end{aligned}
\ee
where $\mathcal{C}_{xz,x}'$ is a slightly deformed  loop of $\mathcal{C}_{xz,x}$, and $\mathcal{D}$ is the surface between these two loops.  The integrand on the right-hand side, however, is a gauge-invariant operator that vanishes on-shell: Using the identification \eqref{eq:A_phi}, the left-hand side reduces to a number, $1$. Similarly, $W_{yz,y}$ also evaluates to one on the boundary. The triviality of both operators implies that the bulk magnetic dipoles mobile in either $xz$- or $yz$- plane become condensed on the boundary. This explains the label $(mm)$ in \eqref{eq:mm}: the gapping term describes a boundary where the magnetic excitations (dipoles) condense in both directions of the two-dimensional boundary.

We summarize the discussions above in table \ref{tab:smooth}. Combining everything, the total ground state degeneracy in the system is 
\be
\log_N \text{GSD}^{(mm)\times (mm)} = l_x+l_y+2l_z-2.
\label{eq:final_smooth}
\ee
Where the two parentheses on the superscript describes the boundary conditions on both toroidal boundaries. 
\begin{table}[htbp]
    \centering
    \begin{tabular}{|c|c|c|}
    \hline
       \vtop{\hbox{\strut Non-commuting}\hbox{\strut \quad operators}}  &  \vtop{\hbox{\strut \quad Copies of}\hbox{\strut Heisenberg alg.}}   & \vtop{\hbox{\strut Contribution}\hbox{\strut \quad to GSD}}\\ \hline
       $\hat{W}^x$, $W_{xz,z}$  &  $l_y$ & \multirow{2}{*}{$N^{l_y+l_z-1}$} \\ \cline{1-2}
       $\hat{W}^x$, $W_{xy,y}$  &  $l_z-1$ & \\ \hline
       $\hat{W}^y$, $W_{yz,z}$  &  $l_x$ & \multirow{2}{*}{$N^{l_x+l_z-1}$} \\ \cline{1-2}
       $\hat{W}^y$, $W_{xy,x}$  &  $l_z-1$ & \\ \hline
       \sout{$\hat{W}^z$}, \boldmath{$W_{yz,y}$}  &  $0$ & \multirow{2}{*}{$N^{0}$} \\ \cline{1-2}
       \sout{$\hat{W}^z$}, \boldmath{$W_{xz,x}$}  &  $0$ & \\ \hline
    \end{tabular}
    \caption{Numbers of nontrivial Wilson operators, copies of $\mathbb{Z}_N$ Heisenberg algebras, and contributions to the ground state degeneracy for the (mm)$\times$(mm)-type boundaries. The slashed operators were present in the case without boundaries, but do not preserve the ground state subspace on $T^2\times I$. The operators in bold face are condensed on the boundary.}
    \label{tab:smooth}
\end{table}

Next, we provide two additional ways of counting the ground state degeneracy, from a lattice point of view. The first way amounts to counting the lattice degrees of freedom, while the second way is analogous to the above discussions in the continuum and is counting different ways of threading fluxes through the system.

\subsection{Counting of lattice DOFs}

In this part, we take the alternative view and count the ground state degeneracy using a lattice Hamiltonian. The boundary Hamiltonian for the $\mathbb{Z}_2$ case was worked out in ref. \cite{PhysRevB.96.195139}, and what we present below is a generalization to $\mathbb{Z}_N$. 

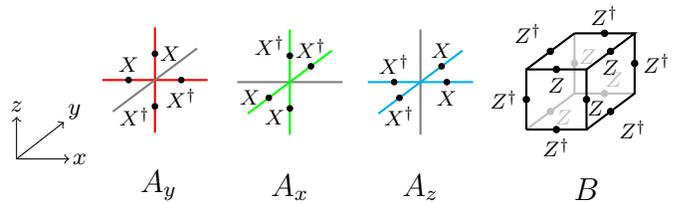
\begin{figure}[H]
\raggedright
\begin{tikzpicture}[scale=0.70]
\hspace{0.1cm}
\draw[color=red, thick] (0,0) -- (1,0);
\draw[color=red, thick] (0,0) -- (-1,0);
\draw[color=red, thick] (0,0) -- (0,-1);
\draw[color=red, thick] (0,0) -- (0,1);
\draw[color=gray, thick] (0,0) -- (0.8,0.6);
\draw[color=gray, thick] (0,0) -- (-0.8,-0.6);
\filldraw 
(0.5,0) circle (1.5pt) node[align=left, below] {\footnotesize $X^\dagger$}
(-0.5,0) circle (1.5pt) node[align=left, above] {\footnotesize $X$}
(0,-0.5) circle (1.5pt) node[align=left, below left=-0.1cm and -0.05cm] {\footnotesize $X^\dagger$}
(0,0.5) circle (1.5pt) node[align=left, right=-0.1cm and -0.05cm] {\footnotesize $X$};
\hspace{-0.3cm}
\filldraw (0.5,-2) node[] {\large $A_y$};
\draw[->] (-2.2,-1.5)--(-1.2,-1.5);
\draw[->] (-2.2,-1.5)--(-1.3,-0.8);
\draw[->] (-2.2,-1.5)--(-2.2,-0.7);
\node at (-1.0,-1.5) {$x$};
\node at (-1.1,-0.6) {$y$};
\node at (-2.2,-0.5) {$z$};
\end{tikzpicture}
\begin{tikzpicture}[scale=0.70]
\hspace{0.3cm}
\draw[color=gray, thick] (0,0) -- (1,0);
\draw[color=gray, thick] (0,0) -- (-1,0);
\draw[color=green, thick] (0,0) -- (0,-1);
\draw[color=green, thick] (0,0) -- (0,1);
\draw[color=green, thick] (0,0) -- (0.8,0.6);
\draw[color=green, thick] (0,0) -- (-0.8,-0.6);
\filldraw 
(0.4,0.3) circle (1.5pt) node[align=left, above] {\footnotesize $X^\dagger$}
(-0.4,-0.3) circle (1.5pt) node[align=left, left] {\footnotesize $X$}
(0,-0.5) circle (1.5pt) node[align=left, below left=-0.1cm and -0.05cm] {\footnotesize $X$}
(0,0.5) circle (1.5pt) node[align=left, above left=-0.1cm and -0.05cm] {\footnotesize $X^\dagger$};
\filldraw (0,-2) node[] {\large $A_x$};
\end{tikzpicture}
\begin{tikzpicture}[scale=0.70]
\hspace{0.5cm}
\draw[color=cyan, thick] (0,0) -- (1,0);
\draw[color=cyan, thick] (0,0) -- (-1,0);
\draw[color=gray, thick] (0,0) -- (0,-1);
\draw[color=gray, thick] (0,0) -- (0,1);
\draw[color=cyan, thick] (0,0) -- (0.8,0.6);
\draw[color=cyan, thick] (0,0) -- (-0.8,-0.6);
\filldraw 
(0.5,0) circle (1.5pt) node[align=left, below] {\footnotesize $X$}
(-0.5,0) circle (1.5pt) node[align=left, above] {\footnotesize $X^\dagger$}
(0.4,0.3) circle (1.5pt) node[align=left, above] {\footnotesize $X$}
(-0.4,-0.3) circle (1.5pt) node[align=left, below] {\footnotesize $X^\dagger$};
\filldraw (0,-2) node[] {\large $A_z$};
\end{tikzpicture}
\begin{tikzpicture}[scale=0.80]
\hspace{0.5cm}
\draw[color=lightgray, thick] (0,0) -- (0.8,0.6) -- (0.8,1.6);
\draw[color=lightgray, thick] (0.8,0.6) -- (1.8,0.6);
\draw[thick] (0,0) -- (0,1) -- (1,1) -- (1,0) -- (0,0);
\draw[thick] (0,1) -- (0.8,1.6);
\draw[thick] (0.8,1.6) -- (1.8,1.6) -- (1,1);
\draw[thick] (1.8,1.6) -- (1.8,0.6) -- (1,0);
\filldraw 
(0.5,0) circle (1.5pt) node[align=left, below] {\footnotesize $Z^\dagger$}
(0,0.5) circle (1.5pt) node[align=left, left] {\footnotesize $Z^\dagger$}
(0.5,1) circle (1.5pt) node[align=left, below] {\footnotesize $Z$}
(1,0.5) circle (1.5pt) node[align=left, below right=-0.1cm and -0.1cm] {\footnotesize $Z$}
(0.4,1.3) circle (1.5pt) node[align=left, above left] {\footnotesize $Z^\dagger$}
(1.4,1.3) circle (1.5pt) node[align=left, below] {\footnotesize $Z$}
(1.3,1.6) circle (1.5pt) node[align=left, above] {\footnotesize $Z^\dagger$}
(1.8,1.1) circle (1.5pt) node[align=left, right] {\footnotesize $Z^\dagger$}
(1.4,0.3) circle (1.5pt) node[align=left, below right] {\footnotesize $Z^\dagger$}; 
\filldraw[color=lightgray](0.4,0.3) circle (1.5pt) node[align=left, below right=-0.14cm and -0.05cm] {\footnotesize $Z$}
(1.3,0.6) circle (1.5pt) node[align=left, below right=-0.4cm and -0.1cm] {\footnotesize $Z$}
(0.8,1.1) circle (1.5pt) node[align=left, below right=-0.3cm and -0.05cm] {\footnotesize $Z$};
\filldraw (1,-1) node[] {\large $B$};
\end{tikzpicture}
\caption{Bulk terms in \eqref{eq:dagger}. $X$ and $Z$ label the generalized Pauli matrices. Each term is a product over all the $X$'s or $Z$'s on the corresponding links. The daggered terms $A_{v,i}^\dagger$'s and $B_c^\dagger$ are straightforward.}
\label{fig:bulk}
\end{figure}

\begin{figure}[htbp]
\raggedright
\begin{tikzpicture}[scale=0.70]\hspace{0.7cm}
\draw[color=red, thick] (0,0) -- (1,0);
\draw[color=red, thick] (0,0) -- (-1,0);
\draw[color=red, thick] (0,0) -- (0,-1);
\draw[color=gray, thick] (0,0) -- (0.8,0.6);
\draw[color=gray, thick] (0,0) -- (-0.8,-0.6);
\filldraw 
(0.5,0) circle (1.5pt) node[align=left, below] {\footnotesize $X^\dagger$}
(-0.5,0) circle (1.5pt) node[align=left, above] {\footnotesize $X$}
(0,-0.5) circle (1.5pt) node[align=left, below left=-0.1cm and -0.05cm] {\footnotesize $X^\dagger$};
\filldraw (0,-2) node[] {\large $A_y$};
\end{tikzpicture}
\begin{tikzpicture}[scale=0.70]\hspace{0.8cm}
\draw[color=gray, thick] (0,0) -- (1,0);
\draw[color=gray, thick] (0,0) -- (-1,0);
\draw[color=green, thick] (0,0) -- (0,-1);
\draw[color=green, thick] (0,0) -- (0.8,0.6);
\draw[color=green, thick] (0,0) -- (-0.8,-0.6);
\filldraw 
(0.4,0.3) circle (1.5pt) node[align=left, above] {\footnotesize $X^\dagger$}
(-0.4,-0.3) circle (1.5pt) node[align=left, left] {\footnotesize $X$}
(0,-0.5) circle (1.5pt) node[align=left, below left=-0.1cm and -0.05cm] {\footnotesize $X$};
\filldraw (0,-2) node[] {\large $A_x$};
\end{tikzpicture}
\begin{tikzpicture}[scale=0.70]\hspace{0.9cm}
\draw[color=cyan, thick] (0,0) -- (1,0);
\draw[color=cyan, thick] (0,0) -- (-1,0);
\draw[color=gray, thick] (0,0) -- (0,-1);
\draw[color=cyan, thick] (0,0) -- (0.8,0.6);
\draw[color=cyan, thick] (0,0) -- (-0.8,-0.6);
\filldraw 
(0.5,0) circle (1.5pt) node[align=left, below] {\footnotesize $X$}
(-0.5,0) circle (1.5pt) node[align=left, above] {\footnotesize $X^\dagger$}
(0.4,0.3) circle (1.5pt) node[align=left, above] {\footnotesize $X$}
(-0.4,-0.3) circle (1.5pt) node[align=left, below] {\footnotesize $X^\dagger$};
\filldraw (0,-2) node[] {\large $A_z$};
\end{tikzpicture}
\begin{tikzpicture}[scale=0.80]
\hspace{1cm}
\draw[color=lightgray, thick] (0,0) -- (0.8,0.6) -- (0.8,1.6);
\draw[color=lightgray, thick] (0.8,0.6) -- (1.8,0.6);
\draw[thick] (0,0) -- (0,1) -- (1,1) -- (1,0) -- (0,0);
\draw[thick] (0,1) -- (0.8,1.6);
\draw[thick] (0.8,1.6) -- (1.8,1.6) -- (1,1);
\draw[thick] (1.8,1.6) -- (1.8,0.6) -- (1,0);
\filldraw 
(0.5,0) circle (1.5pt) node[align=left, below] {\footnotesize $Z^\dagger$}
(0,0.5) circle (1.5pt) node[align=left, left] {\footnotesize $Z^\dagger$}
(0.5,1) circle (1.5pt) node[align=left, below] {\footnotesize $Z$}
(1,0.5) circle (1.5pt) node[align=left, below right=-0.1cm and -0.1cm] {\footnotesize $Z$}
(0.4,1.3) circle (1.5pt) node[align=left, above left] {\footnotesize $Z^\dagger$}
(1.4,1.3) circle (1.5pt) node[align=left, below] {\footnotesize $Z$}
(1.3,1.6) circle (1.5pt) node[align=left, above] {\footnotesize $Z^\dagger$}
(1.8,1.1) circle (1.5pt) node[align=left, right] {\footnotesize $Z^\dagger$}
(1.4,0.3) circle (1.5pt) node[align=left, below right] {\footnotesize $Z^\dagger$}; 
\filldraw[color=lightgray](0.4,0.3) circle (1.5pt) node[align=left, below right=-0.14cm and -0.05cm] {\footnotesize $Z$}
(1.3,0.6) circle (1.5pt) node[align=left, below right=-0.4cm and -0.1cm] {\footnotesize $Z$}
(0.8,1.1) circle (1.5pt) node[align=left, below right=-0.3cm and -0.05cm] {\footnotesize $Z$};
\filldraw (1,-1) node[] {\large $B$};
\end{tikzpicture}
\caption{(mm)-type or smooth boundary Hamiltonian terms. $A_x$, $A_y$, $A_z$ shown here are on the top boundary of $T^2\times I$. The terms on the bottom boundary is analogous. 
Notice that $A_{v,i}$'s here result from directly erasing all top legs in Fig.\ref{fig:bulk}.}
\label{fig:mm}
\end{figure}
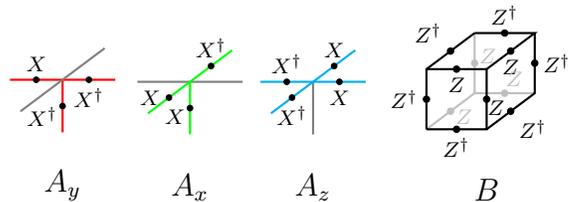

The system is defined on a three-dimensional cubic lattice and the degrees of freedom live on the links of this lattice. The Hamiltonian is a summation over all the terms shown in figure \ref{fig:bulk} in the bulk, and figure \ref{fig:mm} on the boundaries,  
\be
\label{eq:dagger}
H=-\sum_{v,i}(A_{v,i}+A^\dagger_{v,i} )-\sum_{c} (B_c+B_c^\dagger)+\text{h.c.},
\ee
where $v$ labels vertices, $c$ labels cubes, and $i=x, y, z$ labels three different types of vertex operators (operators that act on the vertices of the lattice, not to be confused with the $V_I$ operators introduced in \eqref{eq:V}). 
$A_{v,i}$ and $B_c$ are defined  via the generalized $(N\times N)$-dimensional Pauli matrices $X$ and $Z$ which satisfy
\begin{equation}
    ZX=\omega XZ, \quad \omega=e^{2\pi i/N},
\end{equation}
and they share the same set of eigenvalues $\{1,\omega,\dots,\omega^{N-1}\}$. All terms in the Hamiltonian commute with each other and can therefore be simultaneously diagonalized. 

The boundary described in figure \ref{fig:mm} is  called the smooth boundary since there is no dangling open tail on the surface. Because of this, the magnetic excitations (dipoles of fractons) in the bulk that are mobile in $xz$- or $yz$-planes, when moved towards the boundary, can freely pass through the boundary and vanish into the vacuum. Alternatively, one can create magnetic dipoles on the boundary by acting $X$ on the surface links. These excitations are thus condensed at the boundary, matching with what we discussed in the continuum. 

We will count the ground state degeneracy on $T^2\times I$ by subtracting the number of independent constraints from the number of degrees of freedom.  This explicitly computes the dimension of the ground state subspace.
Recall that the number of sites in the three directions are $l_x,$ $l_y$ and $l_z$, respectively. For a $T^2\times I$ topology with two toroidal boundaries perpendicular to the $z$-axis, the total number of degrees of freedom is the total number of links in the system: $3l_xl_yl_z-l_xl_y$. Each term in the Hamiltonian contributes one constraint to the system, giving $2\times [3l_xl_yl_z+(l_z-1)l_xl_y]$ constraints in total, not necessarily independent from each other. The factor of $2$ comes from the coexistence of daggered and undaggered terms. 

Among those constraints, there are relations to consider. First of all, the daggered and undaggered Hamiltonian terms are simply related by Hermitian conjugation, so only half of them are independent. We in the following therefore only consider $A_{v,i}$ and $B_c$ in \eqref{eq:dagger}.  Secondly, $A_{v,x} A_{v,y} A_{v,z}=1$ on each vertex $v$, giving $l_xl_yl_z$ relations. This is because upon acting $A_{v,x} A_{v,y} A_{v,z}$, each link around $v$ is acted on by both $X$ and $X^\dagger$ operators. Thirdly, the product of all $A_{v,i}$'s in each plane perpendicular to the $i$-direction is always $1$, so we end up with $(l_x+l_y+l_z)$ additional relations. However, the above mentioned two types of relations are again not independent: the product of all $A_{v,x}$'s throughout the whole system is trivial, because this product can be decomposed into products of $A_{v,x}$'s within each $yz$-plane, and then further multiplied over all planes. Similarly, the product of all $A_{v,y}$'s or $A_{v,z}$'s throughout the whole system is also trivial. We can combine these into the following expression 
\be
\bigg(\prod_v A_{v,x}\bigg) \bigg(\prod_{v'} A_{v',y}\bigg) \bigg(\prod_{v''} A_{v'',z}\bigg) = 1.
\ee
However, the left-hand side can be easily rewritten as
\be
\prod_v A_{v,x} A_{v,y} A_{v,z} =1,
\ee
where the equality can be easily seen from $A_{v,x} A_{v,y} A_{v,z}=1$. So we now have  $l_xl_yl_z+(l_x+l_y+l_z)-1$ number of \textit{independent} relations. Finally, the product of all $B_c$'s in each $xy$-plane is always $1$ ($Z$ on the opposite edges on each face of the cube are always Hermitian conjugate of each other), and we have altogether $l_z-1$ of such planes, so they contribute another set of relations. Taking into account of everything, we have
\be
\begin{split}
\log_N \text{GSD}= & (\text{\# d.o.f.})-(\text{\# constraints}) \\
& + (\text{\# independent relations}).
\end{split}
\label{eq:constraints}
\ee
and therefore arrive at \eqref{eq:final_smooth} again, matching perfectly with the countings in the continuum.

\subsection{Counting of string operators}

Next we turn to the counting of different ways of threading fluxes, or equivalently the counting of different string operators. In quantum information terms, this is also the counting of logical operators.

\begin{figure}[htbp]
    \centering
    \includegraphics[scale=0.25]{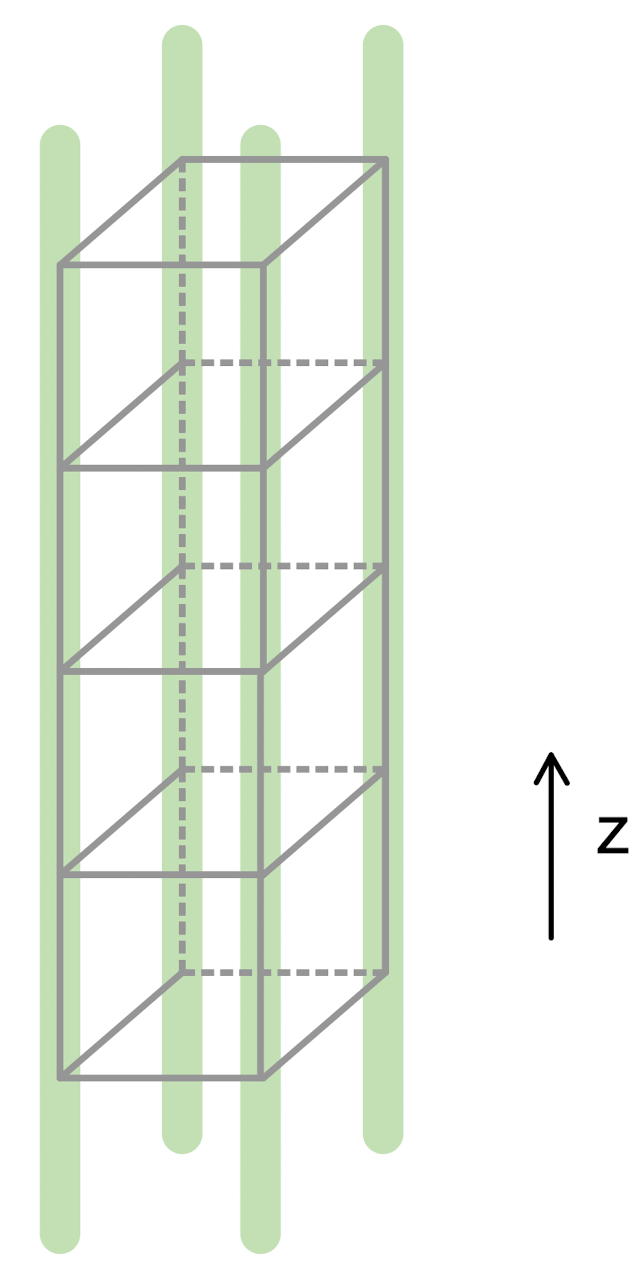}\\
    \vskip 0.5cm
    \includegraphics[scale=0.4]{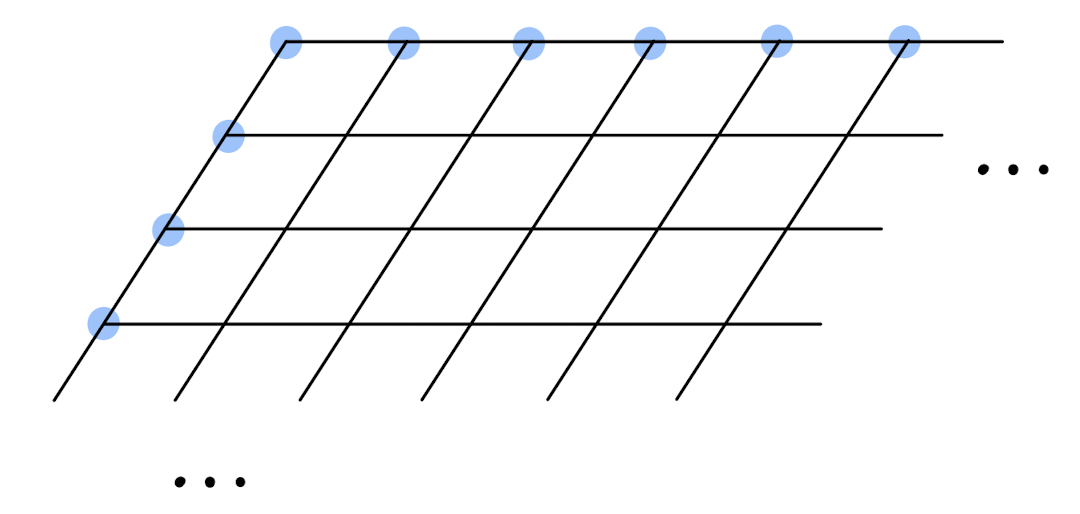}
    \caption{Topology here is $T^3$. Top: threading four $z$-fluxes at the vertices surrounding a plaquette in each $xy$-plane is equal to a product of cube operators along the $z$-loop. Bottom: Knowing the $z$-fluxes threading through the blue vertices is sufficient to derive the fluxes threading the remaining vertices.}
    \label{fig:electric_T3}
\end{figure}

Let us for a moment go back to the case without boundaries, namely, fix the topology to be $T^3$. Starting from the ground state, we consider creating a dipole of $z$-lineons (mobile along the $z$-direction), then winding one of them around the non-contractible $z$-loop, and finally annihilating the dipole. This can be achieved by a string operator which is the product of the (generalized) Pauli operator $Z$'s on the $z$-links around this $z$-loop. Since lineons correspond to the violations of the vertex $A_v$ terms, we call such a procedure the ``$z$-threading'' of electric flux. We can choose to thread electric flux along $z$ at any $(x,y)$ coordinates. Therefore na\"ively, there are $l_x l_y$ ways of doing $z$-threading. However, they are not all independent of each other. As shown in figure \ref{fig:electric_T3}, for any plaquette on a fixed $xy$-plane, threading electric $z$-fluxes at all four corners of this plaquette together is trivial, because this just amounts to multiplying all cube $B$ operators along the $z$-loop. From this, one can easily derive that there are only $l_x+l_y-1$ independent electric string operators along $z$ passing through this plane. 
Since we have three different directions, altogether there are $2l_x+2l_y+2l_z-3$ ways of threading electric fluxes when there is no boundary. 

Now we return to the $T^2\times I$ case when the $z$-direction is open, with both boundaries being smooth. The string operator can no longer wind around the full $z$-loop and can only start/end at the bottom/top boundaries. Such string operators comprising $Z$ operators, however, will not commute with the boundary $A_x$ or $A_y$ terms. 
As in the $T^3$ case, we can still count the independent string operators along the remaining two directions, giving once again \eqref{eq:final_smooth}. 

\section{Rough \texorpdfstring{\MakeLowercase{(ee)}}{}  boundaries}
\label{sec:ee_ee}

In this section, we focus on the case where the following gapping term is added on each independent boundary, 
\be
\L_g^{(ee)}=g_x \cos(N\hat{\vphi}^{x})+g_y \cos(N\hat{\vphi}^{y}),
\label{eq:ee}
\ee
which at large $g_i$, leads to $\hat{\vphi}^{i}=2\pi \hat{m}^i/N$, with $\hat{m}^i(x,y)\in \Z$.  One can again check that the spectrum is gapped, 
\be
\omega^2=-\frac{2\pi V_{11} g_x g_y}{g_x+g_y}+ \frac{k_x^2k_y^2}{4N^2}(4V_{11}V_{22}-(V_{12}+V_{21})^2),
\ee
with the gap proportional to $\sqrt{g_x g_y/(g_x+g_y)}.$ We will soon see that the electric excitations are condensed on this boundary, justifying the name $(ee)$ in \eqref{eq:ee}.

We turn to the $T^2\times I$ geometry. Again we have $l_z$ number of sites along the $z$-direction, and $l_z-1$ number of links excluding the open tails. The $\hat{W}^x(y,z)$ operators, namely $x$-lineons,  when evaluated on the boundary, have definite values because of the gapping terms \eqref{eq:ee}, but remain nontrivial in the bulk.  This reflects the condensation of the  electric planon/dipole operators $\hat{W}^{xz}$ on the boundary:
\be
 \hat{W}^{xz,z}(y_1,y_2,\hat{\C}_{xz,z}\mid) 
=  \exp \left[i\int_{y_1}^{y_2}dy \int_{z_b}^{z_t} dz\ \p_y \p_z \hat{\vphi}^x \right],
\label{eq:e_condense}
\ee
where $\hat{\C}_{xz,z}\mid$ is an open path living in the $xz$-plane and connecting the top and bottom boundaries. The integral thus reduces to two separated $y$-integrals on the two boundaries, which evaluate to constants when $g_x$ and $g_y$ are large. 
Henceforth, we say the dipoles of electric $x$-lineons are \textit{condensed} on the boundary. This justifies the name of $(ee)$-type boundary. Similarly, $y$-lineons $\hat{W}^y(x,z)$ also evaluates to numbers on the boundaries, corresponding to the condensation of $\hat{W}^{yz}$ operators on the boundaries. In this in contrast to the smooth boundary case where the magnetic planons are condensed on the boundaries. Finally, the $\hat{W}^z$ operators can still start/end on  the top/bottom boundaries although the $z$-loop is broken, and their expectation values are  $\Z_N$ phases.

Turning to the unhatted magnetic operators, $W_{xz,z}$ and $W_{yz,z}$ create boundary excitations and therefore should not be included when analyzing the ground state subspace.
There are $l_z$  of $W_{xy,y}(z_1,z_2)$ which form $l_z$ copies of Heisenberg algebra with $\hat{W}^x(y,z)$ when $z_1<z<z_2$. Similarly, $W_{xy,x}(z_1,z_2)$ and $\hat{W}^y(x,z)$ form another $l_z$ copies of Heisenberg algebra. Finally, $W_{yz,y}(x_1,x_2)$,  $W_{xz,x}(y_1,y_2)$ and $\hat{W}^z(x,y)$ together form $l_x+l_y-1$ independent copies of Heisenberg algebras, due to the constraint discussed in \eqref{eq:W_constraint}.

We summarize the discussions above in table \ref{tab:rough}. The total ground state degeneracy in this case of two rough boundaries is then,
\be
\log_N \text{GSD}^{(ee)\times (ee)} = l_x+l_y+2l_z-1.
\label{eq:final_rough}
\ee
\begin{table}[htbp]
    \centering
    \begin{tabular}{|c|c|c|}
    \hline
       \vtop{\hbox{\strut Non-commuting}\hbox{\strut \quad operators}}  &  \vtop{\hbox{\strut \quad Copies of}\hbox{\strut Heisenberg alg.}}   & \vtop{\hbox{\strut Contribution}\hbox{\strut \quad to GSD}}\\ \hline
       {\boldmath {$\hat{W}^x$}}, \sout{$W_{xz,z}$} &  $0$ & \multirow{2}{*}{$N^{l_z}$} \\ \cline{1-2}
       $\hat{W}^x$, $W_{xy,y}$  &  $l_z$ & \\ \hline
       {\boldmath{$\hat{W}^y$}}, \sout{$W_{yz,z}$}  &  $0$ & \multirow{2}{*}{$N^{l_z}$} \\ \cline{1-2}
       $\hat{W}^y$, $W_{xy,x}$  &  $l_z$ & \\ \hline
       $\hat{W}^z$, $W_{yz,y}$  &  $l_x$ & \multirow{2}{*}{$N^{l_x+l_y-1}$} \\ \cline{1-2}
       $\hat{W}^z$, $W_{xz,x}$  &  $l_y$ & \\ \hline
    \end{tabular}
    \caption{Rough boundaries. The slashed-out operators do not preserve the ground state subspace. The operators in boldface are condensed on the boundaries.}
    \label{tab:rough}
\end{table}

\subsection{Counting of lattice DOF}

\begin{figure}[htbp]
\raggedright
\begin{tikzpicture}[scale=0.70]\hspace{0.7cm}
\draw[color=red, thick] (0,0) -- (1,0);
\draw[color=red, thick] (0,0) -- (-1,0);
\draw[color=red, thick] (0,0) -- (0,-1);
\draw[color=red, thick] (0,0) -- (0,1);
\draw[color=gray, thick] (0,0) -- (0.8,0.6);
\draw[color=gray, thick] (0,0) -- (-0.8,-0.6);
\filldraw 
(0.5,0) circle (1.5pt) node[align=left, below] {\footnotesize $X^\dagger$}
(-0.5,0) circle (1.5pt) node[align=left, above] {\footnotesize $X$}
(0,-0.5) circle (1.5pt) node[align=left, below left=-0.1cm and -0.05cm] {\footnotesize $X^\dagger$}
(0,0.5) circle (1.5pt) node[align=left, right] {\footnotesize $X$};;
\filldraw (0,-2) node[] {\large $A_y$};
\end{tikzpicture}
\begin{tikzpicture}[scale=0.70]\hspace{0.8cm}
\draw[color=gray, thick] (0,0) -- (1,0);
\draw[color=gray, thick] (0,0) -- (-1,0);
\draw[color=green, thick] (0,0) -- (0,1);
\draw[color=green, thick] (0,0) -- (0,-1);
\draw[color=green, thick] (0,0) -- (0.8,0.6);
\draw[color=green, thick] (0,0) -- (-0.8,-0.6);
\filldraw 
(0.4,0.3) circle (1.5pt) node[align=left, above] {\footnotesize $X^\dagger$}
(-0.4,-0.3) circle (1.5pt) node[align=left, left] {\footnotesize $X$}
(0,-0.5) circle (1.5pt) node[align=left, below left=-0.1cm and -0.05cm] {\footnotesize $X$}
(0,0.5) circle (1.5pt) node[align=left, left] {\footnotesize $X^\dagger$};
\filldraw (0,-2) node[] {\large $A_x$};
\end{tikzpicture}
\begin{tikzpicture}[scale=0.70]\hspace{0.9cm}
\draw[color=cyan, thick] (0,0) -- (1,0);
\draw[color=cyan, thick] (0,0) -- (-1,0);
\draw[color=gray, thick] (0,0) -- (0,-1);
\draw[color=gray, thick] (0,0) -- (0,1);
\draw[color=cyan, thick] (0,0) -- (0.8,0.6);
\draw[color=cyan, thick] (0,0) -- (-0.8,-0.6);
\filldraw 
(0.5,0) circle (1.5pt) node[align=left, below] {\footnotesize $X$}
(-0.5,0) circle (1.5pt) node[align=left, above] {\footnotesize $X^\dagger$}
(0.4,0.3) circle (1.5pt) node[align=left, above] {\footnotesize $X$}
(-0.4,-0.3) circle (1.5pt) node[align=left, below] {\footnotesize $X^\dagger$};
\filldraw (0,-2) node[] {\large $A_z$};
\end{tikzpicture}
\begin{tikzpicture}[scale=0.80]
\hspace{1cm}
\draw[thick] (0,0) -- (0.8,0.6) -- (0.8,1.6);
\draw[thick] (0.8,0.6) -- (1.8,0.6);
\draw[thick] (0,0) -- (0,1);
\draw[thick] (0,0) -- (1,0) -- (1,1);
\draw[thick] (1.8,1.6) -- (1.8,0.6) -- (1,0);
\filldraw 
(0.5,0) circle (1.5pt) node[align=left, below] {\footnotesize $Z^\dagger$}
(0,0.5) circle (1.5pt) node[align=left, left] {\footnotesize $Z^\dagger$}
(1,0.5) circle (1.5pt) node[align=left, below right=-0.1cm and -0.1cm] {\footnotesize $Z$}
(1.8,1.1) circle (1.5pt) node[align=left, right] {\footnotesize $Z^\dagger$}
(1.4,0.3) circle (1.5pt) node[align=left, below right] {\footnotesize $Z^\dagger$}
(0.4,0.3) circle (1.5pt) node[align=left, below right=-0.14cm and -0.05cm] {\footnotesize $Z$}
(1.3,0.6) circle (1.5pt) node[align=left, below right=-0.4cm and -0.1cm] {\footnotesize $Z$}
(0.8,1.1) circle (1.5pt) node[align=left, left] {\footnotesize $Z$};
\filldraw (1,-1) node[] {\large $B$};
\end{tikzpicture}
\caption{(ee)-type or rough boundary Hamiltonian terms.  The vertex operators are the same as in the bulk, while the cube operators are missing the horizontal links on the boundary. }
\label{fig:ee}
\end{figure}
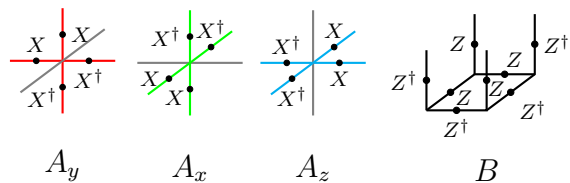

Below we recalculate the ground state degeneracy of the  $(ee)\times(ee)$-type boundary from a lattice perspective as a consistency check. The bulk Hamiltonian was reviewed in figure \ref{fig:bulk}, while the boundary Hamiltonian is shown in figure \ref{fig:ee}. Such a boundary is called a rough one because of the dangling tails on the surface. Due to these open tails, the electric planons are condensed at the boundary, i.e., can freely vanish/emerge, matching with the discussions in the continuum. 
As usual, $l_i$ is the number of sites in the $i$-direction. Since in the $z$-direction there are now dangling tails on the boundary, the number of degrees of freedom is then  $3l_xl_yl_z+l_xl_y$, and the number of constraints, i.e., terms in the Hamiltonian is $2\times (4l_xl_yl_z+l_xl_y)$. 

Next we count the relations among these constraints. First of all, the $A_{v,i}$ and $A_{v,i}^\dagger$ are related by Hermitian conjugate, and similarly for $B_c$ and $B_c^\dagger$. So at most only half of the constraints are independent. Secondly, the product of $A_{x}A_yA_z$ at each vertex is trivial, giving rise to another $l_xl_yl_z$ relations. Thirdly, the product of $A_z$ operators in each $xy$-plane is trivial, giving rise to $l_z$ relations. Furthermore the product of all the $B$ operators in any plane is also trivial, leading to $l_x+l_y+(l_z+1)$ relations. But these relations involving $B$ are not all independent, i.e., there are ``relations among relations'': 
\be
\prod_c B_c=\prod_{\text{different } ij-\text{planes}}\  \prod_{c\ \in\ \text{fixed }\\ ij-\text{plane}} B_c=1,
\ee
the above equation holds for three different combinations of $i, j$, giving rise to two ``relations among relations''. Consequently, we have altogether $l_x+l_y+2l_z-1$ independent relations. Using \eqref{eq:constraints}, we arrive at the ground state degeneracy \eqref{eq:final_rough}.

\subsection{Counting of string operators}

Now we turn to the counting of different flux threadings or independent string operators from a lattice perspective. Since we expect certain electric fluxes to condense, we will count the magnetic fluxes instead. 

\begin{figure}[htbp]
    \centering
    \includegraphics[trim={0 3cm 10cm 2cm}, clip, scale=0.4]{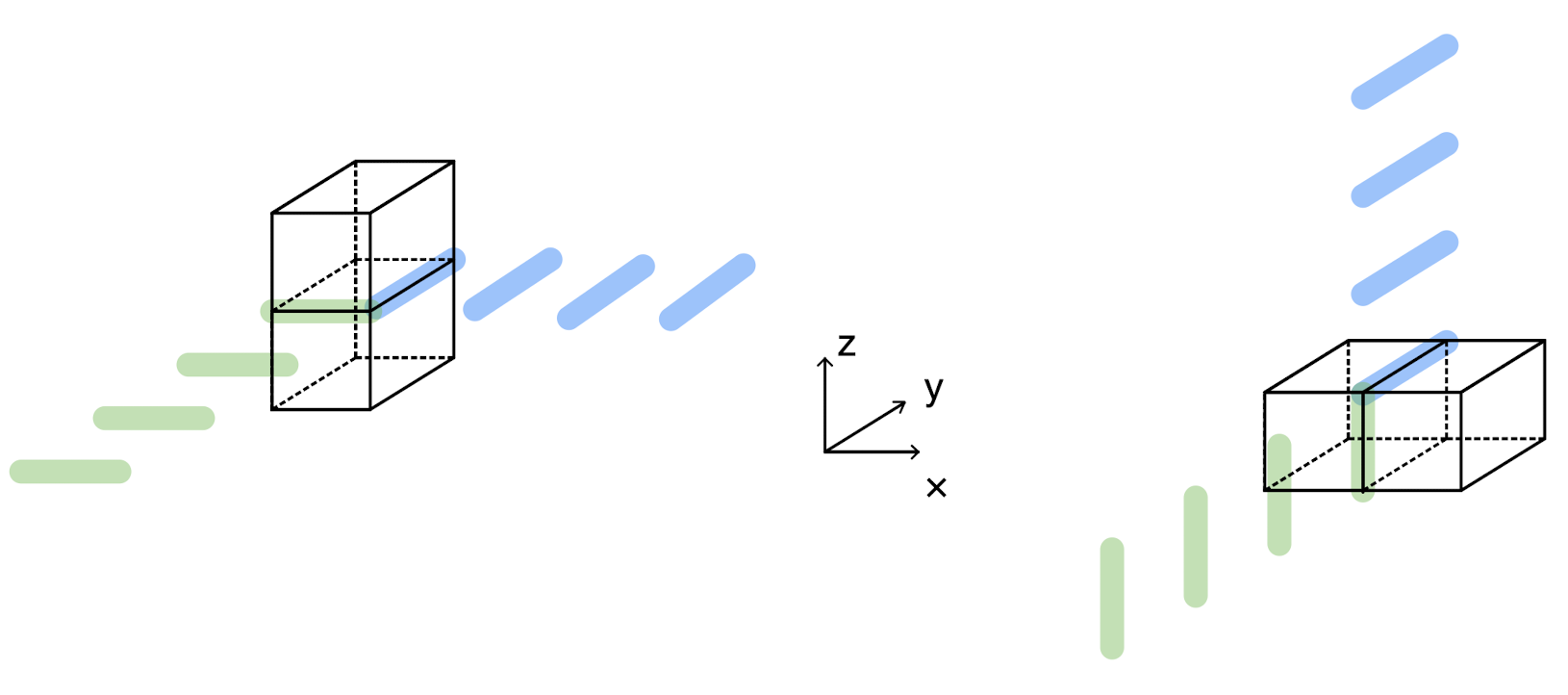}
    \caption{A dipole of fractons separated in the $z$-direction can move freely in the $xy$-plane. It moves along $y$ by acting generalized Pauli $X$ on the green links, while it moves along $x$ by acting $X$ on the blue links. 
    }
    \label{fig:magnetic_T3_1a}
\end{figure}

\begin{figure}[htbp]
    \centering
        \includegraphics[scale=0.4]{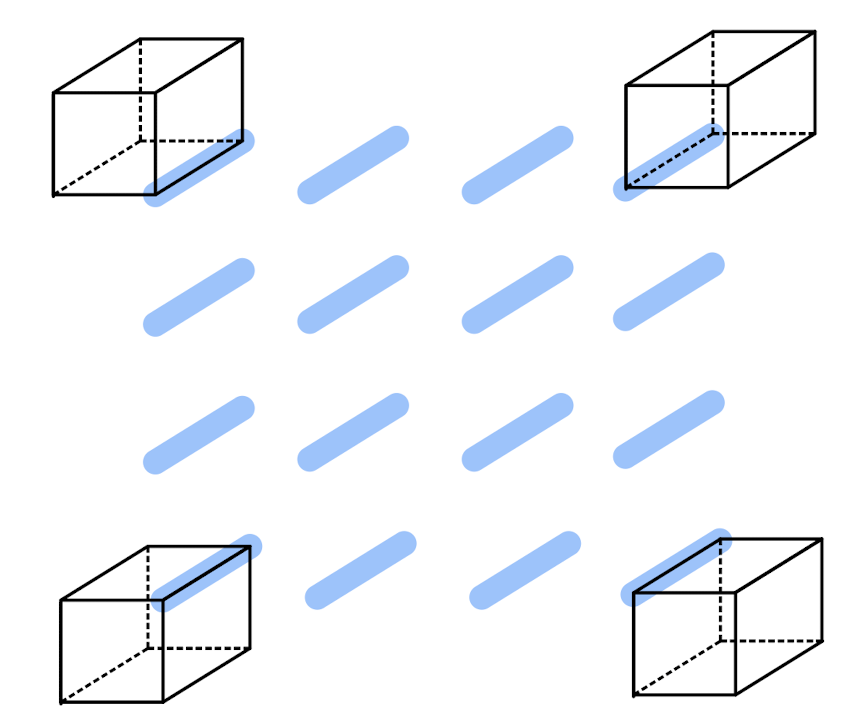}
    \caption{This quadrupole of fractons can be viewed as either two horizontal dipoles separated vertically, or two vertical dipoles separated horizontally. See main text.}
    \label{fig:magnetic_T3_2}
\end{figure}

\begin{figure}[htbp]
    \centering
    \includegraphics[trim={14cm 0.5cm 0 0cm}, clip, scale=0.4]{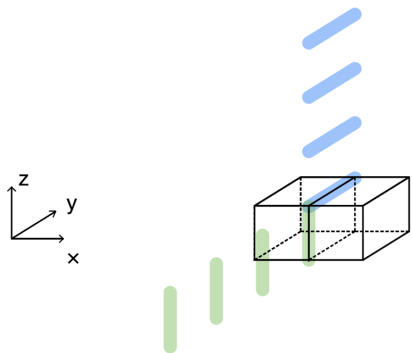}
    \caption{A dipole of fractons separated in the $x$-direction can move freely in the $yz$-plane.}
    \label{fig:magnetic_T3_1b}
\end{figure}

We start with briefly reviewing the counting in the case without boundaries. Consider a dipole of fractons (violated cube terms) separated in the $z$-direction, namely a $z$-lineon. Recall this dipole can move freely in the $xy$-plane, see figure \ref{fig:magnetic_T3_1a}. Therefore, upon acting a string of (generalized) Pauli $X$ operators on all the blue (or green) links, one can thread magnetic fluxes in the $x$- (or $y$-) direction in this $xy$-plane. Since there are $l_z$ number of such $xy$-planes, altogether we have $2l_z$  independent magnetic string-like operators winding around $x$ and $y$ directions.
There are also $l_x$ number of $yz$-planes and $l_y$ number of $xz$-planes, so one would expect that they give altogether $2l_x+2l_y+2l_z$ number of string operators. But these are not all independent: Figure \ref{fig:magnetic_T3_2} shows a quadruple of fractons in the $xz$-plane, which can be split into two dipoles in different ways. It can be viewed either as a dipole of fractons  separated in the $x$-direction (the bottom pair of cubes) and winding around the $z$-direction, 
or as a dipole of fractons separated in the $z$-direction (the left pair of cubes) and winding around the $x$-direction. 
In other words, the blue links in fig. \ref{fig:magnetic_T3_2} come either from the product of the blue links fig. \ref{fig:magnetic_T3_1a} along the $z$-direction, or from the product of blue links of fig. \ref{fig:magnetic_T3_1b} along the $x$-direction. Consequently, each pair of directions gives rise to one global constraint, and we are left with altogether $2l_x+2l_y+2l_z-3$ ways of threading magnetic fluxes in this closed system.

When the $z$-direction is open and the boundaries are rough, dipoles of fractons can no longer wind around the full $z$-loop. The string of Pauli $X$ operators can still start/end at the bottom/top boundaries, but will not commute with the $B$ terms on the boundary. Namely, we will for example lose the string operator associated with the blue links in fig. \ref{fig:magnetic_T3_1b}.  In addition, the global constraints involving string operators in the $z$-direction will also be lost, and there are two of these. Combining everything, the ground state degeneracy thus reduces to \eqref{eq:final_rough} again.

\section{Mixed smooth-rough boundaries}
\label{sec:ee_mm}

In the last two sections, we have discussed the cases where the two gapped boundaries in the system are of the same type, i.e., either both smooth, or both rough. It can also happen that one of the two boundaries is smooth, and the other is rough, such that we have a $(mm)\times (ee)$ or $(ee)\times (mm)$ type of boundaries. The result amounts to the following table \ref{tab:mixed}, where we slash out all the operators that have already been slashed out in either table \ref{tab:smooth} or table \ref{tab:rough}, and boldface all the operators that were bold in either tables.
\begin{table}[htbp]
    \centering
    \begin{tabular}{|c|c|c|}
    \hline
       \vtop{\hbox{\strut Non-commuting}\hbox{\strut \quad operators}}  &  \vtop{\hbox{\strut \quad Copies of}\hbox{\strut Heisenberg alg.}}   & \vtop{\hbox{\strut Contribution}\hbox{\strut \quad to GSD}}\\ \hline
       {\boldmath {$\hat{W}^x$}}, \sout{$W_{xz,z}$} &  $0$ & \multirow{2}{*}{$N^{l_z}$} \\ \cline{1-2}
       $\hat{W}^x$, $W_{xy,y}$  &  $l_z$ & \\ \hline
       {\boldmath{$\hat{W}^y$}}, \sout{$W_{yz,z}$}  &  $0$ & \multirow{2}{*}{$N^{l_z}$} \\ \cline{1-2}
       $\hat{W}^y$, $W_{xy,x}$  &  $l_z$ & \\ \hline
       \sout{$\hat{W}^z$}, \boldmath{$W_{yz,y}$}  &  $0$ & \multirow{2}{*}{$N^{0}$} \\ \cline{1-2}
       \sout{$\hat{W}^z$}, \boldmath{$W_{xz,x}$}  &  $0$ & \\ \hline
    \end{tabular}
    \caption{Mixed boundary $(mm)\times (ee)$. The slashed operators do not preserve the ground state subspace. The operators in bold face are condensed on at least one boundary.}
    \label{tab:mixed}
\end{table}
The ground state degeneracy is
\be 
\log_N GSD^{(ee)\times (mm)}=2l_z.
\label{eq:final_mixed}
\ee

The counting of lattice degrees of freedom goes as follows. The number of links is $3l_xl_yl_z$, and the number of constraints is $4l_xl_yl_z$. The relations among the constraints include again $A_xA_yA_z=1$ at each vertex, leading to $l_xl_yl_z$ number of relations. The product of $A_z$ in each $xy$-plane is trivial, and the product of $B$ in each $xy$-plane is also trivial, giving rise to additional $2l_z$ relations. There are no ``relations among relations''. So the final result is again \eqref{eq:final_mixed}.

\section{Anisotropic \texorpdfstring{\MakeLowercase{(me)}}{} boundaries}
\label{sec:em_em}

In this section, we consider gapped boundaries that break the fourfold rotational symmetry and condense the magnetic (electric) planons only in the $x$- ($y$-) direction, respectively.

We first add the auxiliary fields $\chi$, $\hat{\chi}$, decoupled from the boundary fields $\vphi$ and $\hat{\vphi}^i$, to the boundary Lagrangian:
\be
\L_{\text{aux}}=-\frac{i}{2\pi}
(\p_y \chi) \p_0 \p_x \hat{\chi}.
\ee
$\L_{\text{aux}}$ attaches to the system the boundary theory of a trivial $\Z_1$ X-cube model, which can be easily gapped out by itself. $\L_{\text{aux}}$ will not modify the ground state degeneracy, quasiparticle contents or any topological property of the system. A single $(me)$ boundary then amounts to further adding the following gapping terms (an $(em)$-type boundary can be easily obtained by exchanging $x\leftrightarrow y$),
\begin{equation}
\begin{split}
    \mathcal{L}_g^{(me)} & =  g_1 \cos (Na\p_x\hat{\vphi}^x+a\p_x\hat{\chi})+g_{2} \cos (N\hat{\vphi}^y)\\
    & \quad +g_{3}\cos (Na\p_y\vphi-Na\p_y\chi).
\end{split}
\label{eq:ani_gap}
\end{equation}
The fact that the UV cutoff $a$ enters the gapping terms is a manifestation of UV/IR mixing. 
The $g_2$ term, when large, pins $\hat{\vphi}^y=2\pi \hat{m}^y/N$, with $\hat{m}^y(x,y)$ generally an integer-valued function of $x$ and $y$. Consequently $\hat{\vphi}^z=-\hat{\vphi}^x-2\pi \hat{m}^y/N$.  
The arrangements of $N$'s inside the cosines of equation \eqref{eq:ani_gap} guarantee that these terms mutually commute with each other, such that they can be simultaneously satisfied without frustration: 
$[N\p_y\vphi-N\p_y\chi,N\p_x\hat{\vphi}^x+\p_x\hat{\chi}]= N^2[\p_y\vphi,\p_x\hat{\vphi}^x]-N[\p_y\chi,\p_x\hat{\chi}]=0.$ Furthermore, the $N$'s guarantee that $\L_g$ implements the condensations of bosonic excitations only. When all the $g_i$'s are large, we have the following relations:
\be
\begin{split}
& \p_x\hat{\vphi}^x=-\p_x\hat{\chi}/N+2\pi \hat{m}^x/N,\quad \hat{\vphi}^y=2\pi \hat{m}^y/N,\\
& \p_y\vphi=\p_y\chi+2\pi m/N,
\end{split}
\label{eq:ani_identifications}
\ee
with both $m$ and $\hat{m}^i$ being spatially dependent, periodic integer-valued functions.  The full boundary Lagrangian, together with the auxiliary term $\L_{\text{aux}}$,  then simply reduces to zero.  

To understand these gapping terms better, we turn to the Wilson operators. We again take two $(me)$ type-boundaries at $z=z_t$ and $z=z_b$, where $\chi(z_b)$ and $\chi(z_t)$ are not necessarily the same, and similarly for $\hat{\chi}(z_b)$ and $\hat{\chi}(z_t).$ We start with the analyses of the hatted electric Wilson operators to see the effects of large $g_1$ and $g_2$. $\hat{W}^x(y,z)$ are still $\Z_N$ operators on the boundaries,
\be
\hat{W}^x\mid=\exp\left[i\oint dx\  (\p_x\hat{\chi}+2\pi\hat{m}^x)/N\right]. 
\ee
Therefore, they give rise to $(l_y+l_z-1)$ independent values as if there is no boundary. 
$\hat{W}^y(x,z)$ simply evaluates to one on the boundary, because of the $g_2$ term and the fact that $\hat{m}^y$ is periodic in $y$. This corresponds to the fact that the electric planons $\hat{W}^{yz}$ are condensed on the boundary, similar to the rough boundary case. But $\hat{W}^y(x,z)$ can still take $(l_z-2)$ different values in the bulk. $\hat{W}^z(x,y)=\exp [i\hat{\vphi}^z\mid^{z_t}_{z_b}]$, creates boundary excitations.  There are, however, additional Wilson operators that connect the top and bottom boundaries, and create no excitations:
\be
\begin{split}
& \hat{W}^{yz,z}(x_1,x_2,\hat{\C}_{yz,z})\\
= & \exp \left[-i\int_{x_1}^{x_2}dx \int_{z_b}^{z_t}dz\ (\p_x\hat{A}^{xy}+\p_z\hat{A}^{yz})\right]\\
= & \exp \left[-i\int_{x_1}^{x_2}dx \int_{z_b}^{z_t}dz\ (\p_x\p_z\hat{\vphi}^{z}+\p_z\p_x\hat{\vphi}^{x})\right].\\
\end{split}
\label{eq:66}
\ee 
It describes a dipole of $y$-lineons separated in the $x$-direction and move between the two boundaries.  Usually this creates excitations on the boundary, but exactly due to the $g_2$ term, we have  $\hat{\vphi}^z=-\hat{\vphi}^x-2\pi\hat{m}^y/N,$ and the expectation value of $\hat{W}^{yz,z}$ is a $\Z_N$ phase for fixed $(x_1,x_2)$. In other words, since the $y$-lineons are condensed on the boundary, $\hat{W}^{yz,z}$ acts within the ground state subspace. They have $(l_x-1)$ independent values, with the ``$-1$'' due to the fact that $\oint dx \p_x \hat{m}^y=0.$

Next we turn to the magnetic Wilson operators. We first look at  $W_{xz,x}(y_1,y_2,\C_{xz,x})$ and $W_{yz,y}(x_1,x_2,\C_{yz,y})$. As discussed near \eqref{eq:quasi-top}, their values are independent of the $z$-coordinate.  Consequently, we take the closed path $\C_{iz,i}$ to span the entire $z$-direction, ending on both boundaries. Then the exponents in both $W_{xz,x}$ and $W_{yz,y}$ are double integrals in $x$ and $y$, with the same integrand, $\p_x\p_y\vphi$, but different integration limits: One is integrated over a non-contractible cycle in $x$, while the other is over $y$. At large $g_1$, 
$\p_x\p_y\vphi$ reduces to $\p_x\p_y\chi+2\pi\p_x m/N$. Both $e^{i\oint dx \p_x\p_y\chi}$ and $e^{i\oint dy \p_x\p_y\chi}$ are trivial since $\L_{\text{aux}}$ has level one. $\oint dx \p_x m=0$ since $m$ is periodic, while $\oint dy \p_x m$ depends on $x$.  Consequently, $W_{xz,x}(y_1,y_2,\C_{xz,x})$ is trivial, reflecting the fact that the magnetic dipoles separated in the $y$-direction are condensed on the boundary, while $W_{yz,y}(x_1,x_2,\C_{yz,y})$ has $l_x$ independent values corresponding to different choices of $(x_1,x_2)$. Among these choices, when $(x_1,x_2)=(0,L_x),$ there is a relation $W_{xz,x}(0,L_y,\C_{xz,x})=W_{yz,y}(0,L_x,\C_{yz,y})$, and the left-hand side is trivial. Hence they give only $(l_x-1)$ number of nontrivial Wilson operators and form Heisenberg algebras with $\hat{W}^{yz,z}$. 

The operators $W_{xz,z}$ are, due to \eqref{eq:ani_identifications}, 
\be
\begin{split}
W_{xz,z}(y_1,y_2,\C_{xz,z}|)
= & \exp\left[i (\chi(y_2)-\chi(y_1))\mid^{z_t}_{z_b}\right.\\
& \left. +2\pi i (m(y_2)-m(y_1)) \mid^{z_t}_{z_b} /N\right].
\end{split}
\ee 
There are $l_y$ number of them, forming Heisenberg algebras with $\hat{W}^x(x,y)$ when $y_1<y<y_2.$  The operators $W_{yz,z}$, on the other hand, are
\be
W_{yz,z}(x_1,x_2,\C_{yz,z}|)=\exp[i\vphi(x_2)-i\vphi(x_1)]^{z_t}_{z_b}.
\ee
They insert excitations on the boundary as usual. 
Finally, the operators $W_{xy}(z_1,z_2,\C_{xy})$ are also nontrivial, 
\be
\begin{split}
& W_{xy,x}(z_1,z_2,\C_{xy,x}) 
= \exp\left[i\int_{z_1}^{z_2} dz\oint_{\mathcal{C}_{xy,x}}dx\ \p_x \p_z \vphi\right],\\
& W_{xy,y}(z_1,z_2,\C_{xy,y}) 
=\exp\left[i\int_{z_1}^{z_2} dz\oint_{\mathcal{C}_{xy,y}}
dy\ \p_y \p_z \vphi\right]. 
\end{split}
\ee
$W_{xy,x}(z_1,z_2,\C_{xy,x})$ and $\hat{W}^y(x,z)$ form $(l_z-2)$ copies of $\mathbb{Z}_N$ Heisenberg algebra, for $z_1<z<z_2$, and similarly for $W_{xy,y}(z_1,z_2,\C_{xy,y})$ and $\hat{W}^x(y,z)$.  In addition, when $(z_1,z_2)=(z_b,z_b+a)$ or $(z_t-a,z_t)$, $W_{xy,y}(z_1,z_2,\C_{xy,y})$ has a nontrivial commutation relation with $\hat{W}^x(y,z_b)$ or $\hat{W}^x(y,z_t)$, giving two additional copies of Heisenberg algebras. Taking into account the global constraint  $W_{xy,y}(0,L_z,\C_{xy,y})=W_{xz,z}(0,L_y,\C_{xz,z|})$, we thus have altogether  $(l_z-2)+(l_z-2+2-1)=2l_z-3$ copies of Heisenberg algebras.

We summarize the above discussions in table \ref{tab:mixed}, and the ground state degeneracy is
\be
\log_N GSD^{(me)\times (me)} = l_x+l_y+2l_z-4.
\label{eq:final_ani}
\ee
\begin{table}[htbp]
    \centering
    \begin{tabular}{|c|c|c|}
    \hline
       \vtop{\hbox{\strut Non-commuting}\hbox{\strut \quad operators}}  &  \vtop{\hbox{\strut \quad Copies of}\hbox{\strut Heisenberg alg.}}   & \vtop{\hbox{\strut Contribution}\hbox{\strut \quad to GSD}}\\ \hline
       $\hat{W}^x$, $W_{xz,z}$ &  $l_y$ & \multirow{2}{*}{$N^{ly+l_z-1}$} \\ \cline{1-2}
       $\hat{W}^x$, $W_{xy,y}$  &  $l_z$ & \\ \hline
       {\boldmath{$\hat{W}^y$}}, \sout{$W_{yz,z}$}  &  $0$ & \multirow{2}{*}{$N^{l_z-2}$} \\ \cline{1-2}
       $\hat{W}^y$, $W_{xy,x}$  &  $l_z-2$ & \\ \hline
       $\hat{W}^{yz,z}$, $W_{yz,y}$  &  $l_x-1$ & \multirow{2}{*}{$N^{l_x-1}$} \\ \cline{1-2}
       \sout{$\hat{W}^z$}, \boldmath{$W_{xz,x}$}  &  $0$ & \\ \hline
    \end{tabular}
    \caption{Ground state degeneracy for the anisotropic boundaries  $(me)\times (me)$. The slashed operators excite the boundary, while the bold operators are condensed on at least one of the two boundaries. }
    \label{tab:ani}
\end{table}

\subsection{Counting using the lattice Hamiltonian}
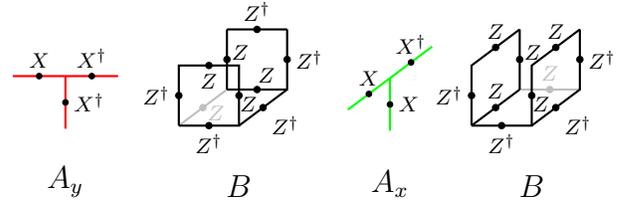
\begin{figure}[htbp]
\raggedright
\begin{tikzpicture}[scale=0.70]
\hspace{0.1cm}
\draw[color=red, thick] (0,0) -- (1,0);
\draw[color=red, thick] (0,0) -- (-1,0);
\draw[color=red, thick] (0,0) -- (0,-1);
\filldraw 
(0.5,0) circle (1.5pt) node[align=left, above] {\footnotesize $X^\dagger$}
(-0.5,0) circle (1.5pt) node[align=left, above] {\footnotesize $X$}
(0,-0.5) circle (1.5pt) node[align=left, right] {\footnotesize $X^\dagger$};
\filldraw (0,-2) node[] {\large $A_y$};
\end{tikzpicture}
\begin{tikzpicture}[scale=0.80]
\hspace{0.2cm}
\draw[color=lightgray, thick] (0,0) -- (0.8,0.6);
\draw[thick] (0.8,0.6) -- (0.8,1.6);
\draw[thick] (0.8,0.6) -- (1.8,0.6);
\draw[thick] (0,0) -- (0,1) -- (1,1) -- (1,0) -- (0,0);
\draw[thick] (0.8,1.6) -- (1.8,1.6);
\draw[thick] (1.8,1.6) -- (1.8,0.6) -- (1,0);
\filldraw 
(0.5,0) circle (1.5pt) node[align=left, below] {\footnotesize $Z^\dagger$}
(0,0.5) circle (1.5pt) node[align=left, left] {\footnotesize $Z^\dagger$}
(0.5,1) circle (1.5pt) node[align=left, below] {\footnotesize $Z$}
(1,0.5) circle (1.5pt) node[align=left, below right=-0.1cm and -0.1cm] {\footnotesize $Z$}
(1.3,1.6) circle (1.5pt) node[align=left, above] {\footnotesize $Z^\dagger$}
(1.8,1.1) circle (1.5pt) node[align=left, right] {\footnotesize $Z^\dagger$}
(1.4,0.3) circle (1.5pt) node[align=left, below right] {\footnotesize $Z^\dagger$}
(1.3,0.6) circle (1.5pt) node[align=left, below right=-0.4cm and -0.1cm] {\footnotesize $Z$}
(0.8,1.1) circle (1.5pt) node[align=left, below right=-0.3cm and -0.05cm] {\footnotesize $Z$}; 
\filldraw[color=lightgray](0.4,0.3) circle (1.5pt) node[align=left, below right=-0.14cm and -0.05cm] {\footnotesize $Z$};
\filldraw (1,-1) node[] {\large $B$};
\end{tikzpicture}
\begin{tikzpicture}[scale=0.70]
\hspace{0.3cm}
\draw[color=green, thick] (0,0) -- (0.8,0.6);
\draw[color=green, thick] (0,0) -- (-0.8,-0.6);
\draw[color=green, thick] (0,0) -- (0,-1);
\filldraw 
(0.4,0.3) circle (1.5pt) node[align=left, above] {\footnotesize $X^\dagger$}
(-0.4,-0.3) circle (1.5pt) node[align=left, above] {\footnotesize $X$}
(0,-0.5) circle (1.5pt) node[align=left, right] {\footnotesize $X$}
;
\filldraw (0,-2) node[] {\large $A_x$};
\end{tikzpicture}
\begin{tikzpicture}[scale=0.80]
\hspace{0.1cm}
\draw[color=lightgray, thick] (0.8,0.6) -- (1.8,0.6);
\draw[thick] (0,0) -- (0.8,0.6);
\draw[thick] (0,0) -- (1,0) -- (1,1) -- (1.8,1.6);
\draw[thick] (0.8,0.6) -- (0.8,1.6);
\draw[thick] (0,0) -- (0,1) -- (0.8,1.6);
\draw[thick] (1.8,1.6) -- (1.8,0.6) -- (1,0);
\filldraw 
(0.5,0) circle (1.5pt) node[align=left, below] {\footnotesize $Z^\dagger$}
(0.4,0.3) circle (1.5pt) node[align=left, above] {\footnotesize $Z$}
(0.4,1.3) circle (1.5pt) node[align=left, above] {\footnotesize $Z$}
(0,0.5) circle (1.5pt) node[align=left, left] {\footnotesize $Z^\dagger$}
(1,0.5) circle (1.5pt) node[align=left, below right=-0.1cm and -0.1cm] {\footnotesize $Z$}
(1.8,1.1) circle (1.5pt) node[align=left, right] {\footnotesize $Z^\dagger$}
(1.4,0.3) circle (1.5pt) node[align=left, below right] {\footnotesize $Z^\dagger$}
(1.4,1.3) circle (1.5pt) node[align=left, above] {\footnotesize $Z$}
(0.8,1.1) circle (1.5pt) node[align=left, below right=-0.3cm and -0.05cm] {\footnotesize $Z$}; 
\filldraw[color=lightgray](1.3,0.6) circle (1.5pt) node[align=left, above] {\footnotesize $Z$};
\filldraw (1,-1) node[] {\large $B$};
\end{tikzpicture}
\caption{$(me)$ and $(em)$ boundary terms. }
\label{fig:em}
\end{figure}

The lattice Hamiltonian on the anisotropic boundary is described in figure \ref{fig:em}, which is smooth along $x$ and rough along $y$. The $\Z_2$ version of this boundary Hamiltonian and the corresponding condensations were discussed in \cite{Daniel}. The $(em)$ boundary case can be obtained by a $90$-degree rotation. 

The total number of degrees of freedom, or links, is  $3l_xl_y(l_z-1)$. The number of constraints which impose the ground-state condition is $2\times[3l_xl_y(l_z-1)+l_xl_y(l_z-1)]$, with the first term coming from $A$ terms at each vertex and the second term coming from the total number of $B$ terms. The Hermitian conjugation relates the daggered and undaggered terms as usual, so we focus only on the undaggered terms. 

There are $l_xl_y(l_z-2)$ relations due to the trivial product of $A_{v,x} A_{v,y} A_{v,z}=1$ at each vertex in the bulk. The trivial product of $A_z$ in each $xy$-plane (excluding the boundary surfaces) gives $(l_z-2)$ relations. The trivial product of $A_y$ over the each $xz$-plane gives $l_y$ relations. Additionally, there are $l_x$ relations coming from the trivial product of all cube $B$ operators in the $yz$-planes, and $(l_z-1)$ relations from the trivial product of all $B$ operators in the $xy$-planes. There is one ``relation among relations'', which is the equality between (a) the product of the cube $B$ operators in each $yz$-plane, and further multiplied over all different $yz$-planes, and (b) the product of the cube $B$ operators in each $xy$-plane, and further multiplied over all different $xy$-planes. Both (a) and (b) are equal to the product of all cube operators in the sample. So altogether we have \eqref{eq:final_ani}.

\subsection{Counting of string operators}

In this section we use the flux-threading argument to count the ground state degeneracy for the anisotropic boundary cases. We can either count the electric or magnetic fluxes or string operators.

\begin{figure}
    \centering
    \includegraphics[scale=0.4]{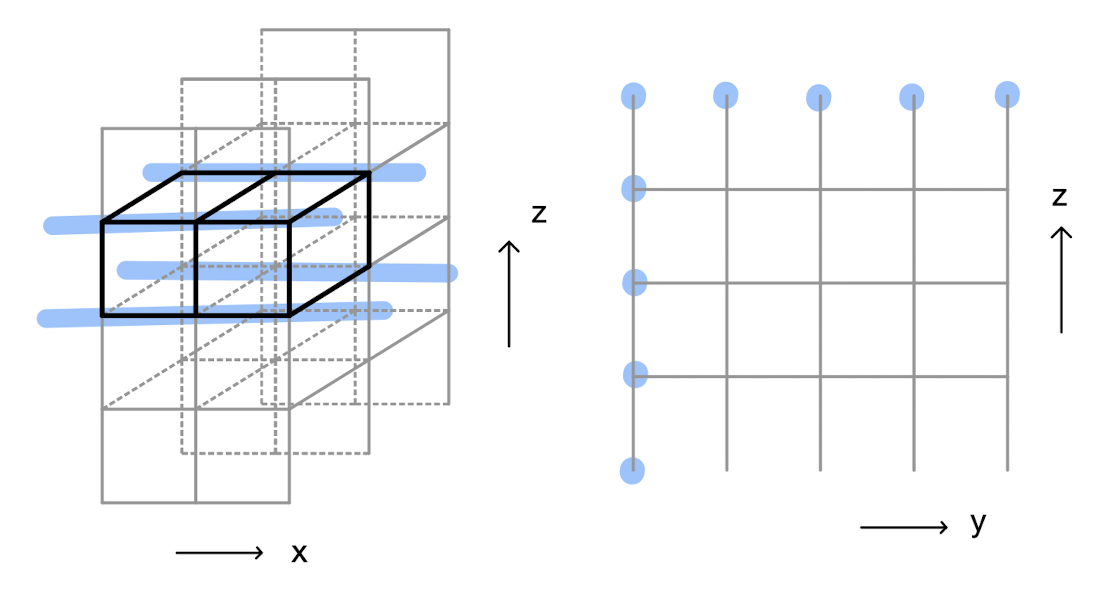}
    \caption{Counting for the electric string operators along $x$-loops are unaffected by boundaries. Left: The relation that the product of four string operators surrounding a plaquette is trivial still holds, because it can be realized by a product of cube operators shown in black. Right: Side view of the geometry. Knowing the fluxes at $l_y+l_z-1$ blue vertices is enough to derive the remaining fluxes.}
    \label{fig:ani_electric_x}
\end{figure}

\begin{figure}[htbp]
    \centering
    \includegraphics[scale=0.4]{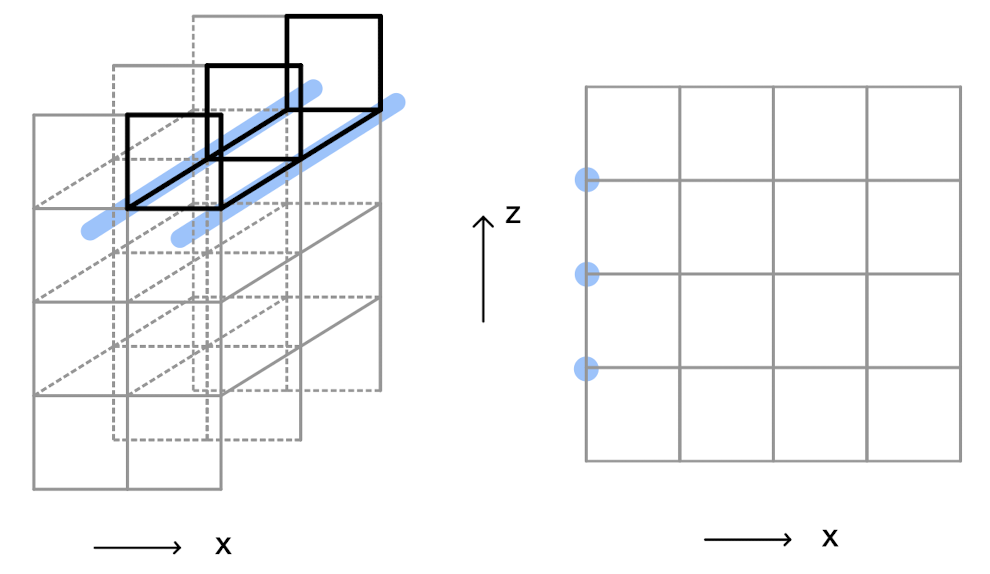}
    \caption{Left: simultaneously threading the two blue $y$-fluxes is trivial because it is a product of black cube operators on the boundary. Right: Front view of the geometry. We only need to know $l_z-2$ number of different fluxes living at the blue vertices in order to derive the rest, since the fluxes are independent of the $x$-coordinate.}
    \label{fig:ani_electric_y}
\end{figure}

We start with counting electric strings of $Z$ operators along the $x$-direction. As shown in figure \ref{fig:ani_electric_x}, this counting is not at all affected by the existence of boundaries, and therefore contributes $l_y+l_z-1$ number of independent strings.
Next, we count the number of independent electric string operators in the $y$-direction. Clearly this is not possible on the boundaries as there are no links along $y$. Consequently, near the boundary, the relation that ``the product of four string operators surrounding a plaquette is trivial'' (as discussed in both figures \ref{fig:electric_T3} and \ref{fig:ani_electric_x} in different directions), reduces to ``the product of two string operators separated in the $x$-direction is trivial''; see the left panel of figure \ref{fig:ani_electric_y}. Consequently, as long as one knows the string operator at one $x$-coordinate, the string operator at other $x$'s can be derived from it. From the right panel of this figure, one see that there are only $(l_z-2)$ independent electric string operators along the $y$-direction.
The strings of $Z$-operators along the $z$-direction, starting and ending on the two boundaries, respectively, no longer commute with all the Hamiltonian terms. However, one can consider the product of Pauli $Z$ operators over a big closed string in the $xz$-plane. The loop spans the entire $z$-direction, but not the $x$-direction. Such operators commute with all bulk and boundary Hamiltonian terms, and correspond to the operators $\hat{W}^{yz,z}$ discussed in \eqref{eq:66}.
There are $(l_x-1)$ of them, with the minus one coming from the fact that when the loop also spans the entire $x$-direction, this string operator is simply a product of the two string operators that wind along $x$ and live on the top and bottom surfaces, which already been counted. Summarizing the counting, we get $l_x+l_y+2l_z-4$ electric string operators in total. 

\begin{figure}
    \centering
     \includegraphics[trim={0 0 17cm 0},clip, scale=0.4]{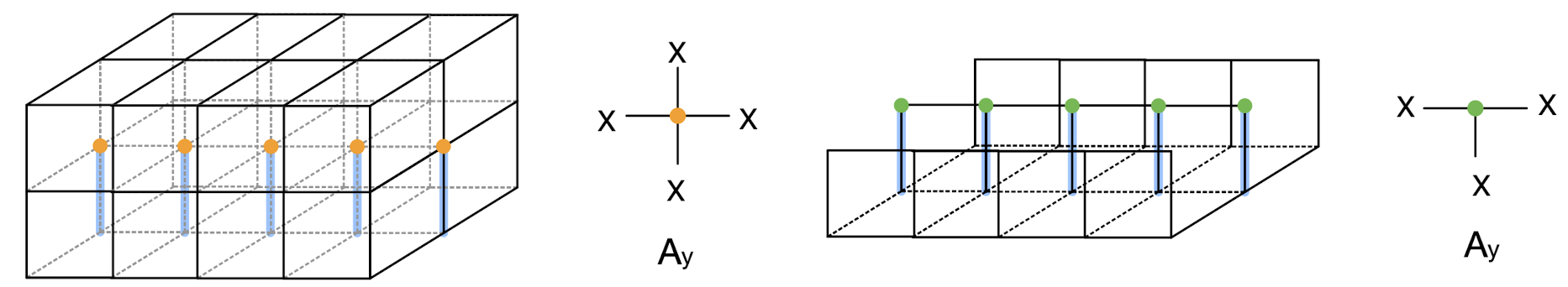}
    \includegraphics[trim={17cm 0 0 0},clip, scale=0.4]{fig_ani_magnetic_x.png}\\
    \caption{A dipole of fractons separated in $y$ can wind along $x$ but gives trivial flux, because this winding procedure can be realized by taking a product of $A_y$ operators in the Hamiltonian. See main text.}
    \label{fig:ani_magnetic_x}
\end{figure}

We can alternatively discuss the magnetic string operators instead and arrive at the same number. Consider a dipole of fractons separated in the $y$-direction. It can still wind along the $x$-direction by a product of Pauli $X$ operators on the blue links in the top panel of figure \ref{fig:ani_magnetic_x}. 
The value of such string operator is independent of the $z$-coordinate, because if one acts the $A_y$ stabilizers on all the orange vertices, the string operator on the bottom layer moves one step up to the top layer. Next, in the bottom panel, if one further acts the $A_y$ stabilizers on all the green vertices on the boundary, the operator moves up and vanishes into the vacuum. Therefore, such magnetic fluxes or string operators are simply trivial. 

We can also consider dipole of fractons separated in the $y$-direction and move it from the bottom to the top boundary. This is realized by a string of Pauli $X$ operators acted on the $x$-links along the $z$-direction. Such string operators still commute with all Hamiltonian terms and there are $l_y$ of them. 

Next we consider the dipole of fractons separated in the $x$-direction. It can still wind along the $y$-direction by acting Pauli $X$ on the green links in figure \ref{fig:magnetic_T3_1b}. There are thus $l_x$ different string operators. (Notice, since the operators $A_x$ are not defined on the boundary, we cannot repeat the argument in figure \ref{fig:ani_magnetic_x}, so each of these string operators is not necessarily trivial.)  The product of these $l_x$ string operators, however, is trivial and reduces to combinations of Hamiltonian terms. So in fact there are only $l_x-1$ independent string operators. A string of Pauli $X$ operators acting on the blue links of figure \ref{fig:magnetic_T3_1b} will not commute with the boundary half-cube terms $B_c$. 

The dipoles of fractons separated in the $z$-direction can wind along both $x$- and $y$-directions as shown back in figure \ref{fig:magnetic_T3_1a}. This gives $2(l_z-2)$ different fluxes. Additionally, a single fracton living on the boundary can freely move along the $y$-direction without forming dipole, by simply acting a string of $X$ operators on the $x$-links on the boundary. Such windings along the $y$-direction are, unlike the fluxes resulting from dipoles of fractons, dependent on the $x$-coordinate now. In other words, there is no product of the Hamiltonian terms that can move the $y$-flux of single fractons on the boundary, in the $x$-direction. But those single boundary fractons with different $x$-coordinates are related through products of planons, which are dipoles of fractons on the boundary, separated in the $x$-direction. Consequently, on each boundary, there is only one independent flux of single fracton that has not been counted. 

There is an additional global constraint among these string operators, corresponding to the fact that the product of all the magnetic $y$-fluxes, in the $z$-direction, is equal to the product of all the magnetic $z$-fluxes in the $y$-direction. 
Summarizing the counting from the magnetic fluxes, we again get $l_x+l_y+2l_z-4$ in total.

\section{General boundary conditions}
\label{sec:Z_N}

As briefly mentioned in Subsection \ref{sec:bdry}, there can be more gapping terms in addition to those introduced in the previous sections. There are roughly two types of them: (1) gapped boundary conditions with general geometries that break translation or/and rotational symmetries; and (2) gapped boundary conditions that preserve the lattice symmetries but condense different quasiparticle contents. We will  discuss each type separately in the two subsections below.

\subsection{Boundaries with domain walls}
\label{subsec:geometry}

In this part, we will discuss multiple boundary conditions living on a single $z=$ constant  boundary. Recall that in the case of a rough boundary, we had
\be
\L_g^{(ee)}=g_x \cos\hat{\vphi}^{x}+g_y \cos\hat{\vphi}^{y},\nonumber
\ee
which at large $g_i$'s pins all $\hat{\vphi}^i$'s to be trivial. Below we will make some modifications on top of this rough boundary condition. The resultant boundary theories can also be viewed as describing domain walls between different elementary boundary conditions.

\subsubsection{Adding one line on the surface}

We start by proceeding just a little step away from the rough boundary, by adding auxiliary fields $\chi$, $\hat{\chi}$ similar to those in the anisotropic boundary case discussed in Section \ref{sec:em_em}, but now only at $y=y_0$: 
\be
\L_{\text{aux}}^{(ee')}=-\frac{i}{2\pi}\delta(y-y_0) (\p_y\chi) (\p_0\p_x\hat{\chi}),
\ee
which again does not modify the contribution to the ground state degeneracy or give rise to nontrivial quasiparticles. The gapping terms are then 
\be
\begin{split}
 \L_g^{(ee')}= & g_1 [1-\delta(y-y_0)/\delta(0)] \cos(N\hat{\vphi}^{x})+g_2 \cos(N\hat{\vphi}^{y})\\
& + g_3 \delta(y-y_0)  \left[\cos(Na\p_y\vphi-Na\p_y\chi)\right.\\
& \left. \ +\cos(Na\p_x\hat{\vphi}^x+a\p_x \hat{\chi})\right].
\end{split}
\label{eq:add_line}
\ee
We have labeled this boundary condition $(ee')$ as it is a minimal modification of the rough boundary condition $(ee)$. When both $g_1$ and $g_2$ are large, $\L_g$ pins $\hat{\vphi}^i$ at all positions on the boundary to be trivial, except at $y=y_0$.  There, the gapping terms are the same as those for the $(me)$ boundary. When all $g_i$'s are large, the full boundary Lagrangian is zero. This boundary condition can be viewed as a nontrivial domain wall between two regions of rough boundaries.

We now count the Wilson operators by examining this system on $T^2\times I$, with both boundaries of the type $(ee')$. Again there are $l_z$ number of sites along the $z$-direction. We start with electric/hatted Wilson operators.  $\hat{W}^x$ operators are trivial on the boundaries except at $y=y_0$, and there are $(l_z-3)$ nontrivial operators in the bulk. $\hat{W}^y$ operators are trivial on the boundaries and  nontrivial only in the bulk. The $\hat{W}^z$ operators evaluate to $\Z_N$ phases except at $y=y_0$, where $\hat{W}^z(x,y_0)$ creates excitations on the boundary. So at $y=y_0$ we need to use $\hat{W}^{yz,z}(x_1,x_2)$ defined in \eqref{eq:66} instead, which gives $l_x-1$ values. The minus one comes from the fact that when  $(x_1,x_2)=(0,L_x)$, the operator is trivial. 
Hence, we get altogether $(l_x-1)+(l_y-1)$ electric Wilson operators that span the entire $z$-direction. 

Next we turn to the magnetic/unhatted Wilson operators $W_{xz,x}(y_1,y_2,\C_{xz,x})$, which are  nontrivial away from $y_0$ when evaluated on the boundaries, and
form $(l_y-1)$ copies of $\mathbb{Z}_N$ Heisenberg algebras with $\hat{W}^z$. On the other hand, $W_{yz,y}(x_1,x_2,\C_{yz,y})$ can be nontrivial and form $l_x$ copies of Heisenberg algebras with $\hat{W}^z$. Between the two sets, there is again one global constraint,  namely $W_{yz,y}(0,L_x,\C_{yz,y})=W_{xz,x}(0,L_y,\C_{xz,x})$, giving us $l_x+l_y-2$ independent copies of Heisenberg algebras.

Wilson operators $W_{yz,z}(x_1,x_2,\C_{yz,z})$ always create boundary excitations and should therefore be excluded. The operators $W_{xz,z}(y_1,y_2,\C_{xz,z})$ create excitations except when $(y_1,y_2)$ sandwich $y_0$, giving a $\Z_N$ phase factor and forming one copy of Heisenberg algebra with $\hat{W}^x$. Finally, $W_{xy,x}(z_1,z_2,\C_{xy,x})$ and $W_{xy,y}(z_1,z_2,\C_{xy,y})$ are both nontrivial in the bulk. In particular, when $z_1\leq z\leq z_2$, they have nontrivial commutation relations with $\hat{W}^y(x,z)$ and $\hat{W}^x(y,z)$, respectively, giving $(l_z-3)+(l_z-1)$ copies of Heisenberg algebra.

Summarizing the countings above, we have
\be
\log_N\ \text{GSD}_{(ee')\times(ee)}=2l_z+l_x+l_y-5.
\ee

\begin{figure}[htbp]
    \centering
    \includegraphics[scale=0.4]{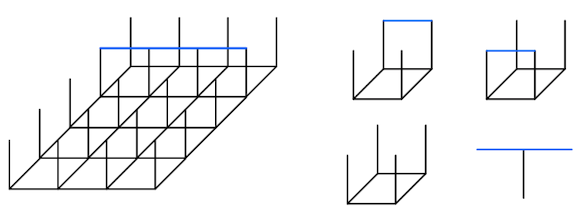}
    \caption{Left: the boundary differs from a rough boundary in figure \ref{fig:ee} in including just one additional blue line along the $x$-direction. Right: the boundary Hamiltonian terms. For convenience, we have omitted the Pauli $X$ and $Z$ operators acting on the links.}
    \label{fig:add_line}
\end{figure}

From the lattice perspective, this boundary condition amounts to adding a single line along the $x$-direction on the boundary surface. The boundary Hamiltonian terms are shown in figure \ref{fig:add_line}, which all mutually commute. There are no additional local terms that commute with these. This is pictorially two regions with rough boundary conditions separated by an anisotropic $(me)$-type boundary condition on the blue line.
Again we put the system on $T^2\times I$, with both boundaries of the $(ee')$ type in figure \ref{fig:add_line}. We now count the magnetic string operators. The electric ones will lead to the same result. 

For the dipoles of fractons separated in the $x$-direction and wind around the $y$-direction, their behavior is unaffected by the boundary and contributes $l_x$ independent magnetic string operators. When dipoles of fractons separated in the $x$-direction move from the one boundary to the other, however, excitations are created on the boundaries. For the dipoles of fractons separated in the $y$-direction and wind around the $x$-direction, most of them remain nontrivial except for the dipole immediately sandwiching the blue line. So they give $(l_y-1)$ independent string operators. In addition, acting $X$ on all the vertical links on the boundary can be interpreted as winding a dipole of fractons separated in the $x$-direction by $L_x$ around the $y$-direction, or winding a dipole of fractons separated in the $y$-direction by $L_y$ around the $x$-direction. So removing this one constraint gives us $(l_x+l_y-2)$ ways of threading magnetic fluxes. 

As for the string operators that are products of $X$ on the $x$-links in the $z$-direction, they correspond to the dipoles of fractons separated in the $y$-direction and moves from the bottom boundary to the top one. Most of the times there are excitations left on the boundaries. Only when the dipole sandwiches the blue line, does the corresponding string operator commute with all Hamiltonian terms and the system remains in the ground state. Consequently, there is only one such string operator to count.

One can also consider dipoles of fractions separated in the $z$-direction and wind around the $x$- or the $y$-direction, giving $2(l_z-2)$ additional independent string operators. Combining all the results above, 
we arrive at $(2l_z+l_x+l_y-5)$ independent windings. 

We would like to briefly comment on the comparison with the boundary theory of the conventional (3+1)d $\mathbb{Z}_N$ toric code \cite{PhysRevB.94.045113}. The (3+1)d $\mathbb{Z}_N$ toric code can also have such a boundary as in the left panel of figure \ref{fig:add_line}. When there is no blue line, magnetic fluxes in the $y$-direction can be nontrivial. But all the magnetic fluxes in the $x$-direction will become completely trivialized due to the existence of this blue line, unlike the X-cube case where only the magnetic fluxes adjacent to the link are trivialized. 

\subsubsection{Smooth and rough boundaries on one surface}

Going one step further ahead from the previous section, now we consider the domain wall between smooth and rough boundaries. The simplest possibility is 
\be
\begin{split}
\L_g^{(ee\backslash mm)} = &\ \Theta(y-y_0) [g_{1x}  \cos (N\hat{\vphi}^x) + g_{1y} \cos (N\hat{\vphi}^y)]\\
& + \delta(y-y_0) g_2 \cos(Na\p_y \vphi -Na\p_y \chi) \\
& + \delta(y-y_0) g_3 \cos(Na\p_x\hat{\vphi}^x + a\p_x\hat{\chi})\\
& +\delta(y-y_0) g_4 \cos (N\hat{\vphi}^y)]\\
& + \Theta(y_0-y) g_5 \cos(N\vphi)\\
\end{split}
\label{eq:ee+mm}
\ee
When $y>y_0$, the rough boundary condition is imposed, and when $y<y_0$, the smooth boundary condition is imposed. We take the definition of the Heaviside step function as $\Theta(y)=1$ when $y>0$, and $\Theta(y)=0$ when $y\leq 0$. At $y=y_0$, the gapping terms are the same as those for the $(ee')$ boundary. The boundary is gapped and the ground state degeneracy can be straightforwardly analyzed using the results from the smooth, rough, and anisotropic boundary conditions.

\subsection{Boundaries with dyon condensations}
\label{subsec:topology}

In this part, we describe gapped boundaries that preserve translation and rotation symmetries, of which the smooth and rough boundaries form a subset. In general, dyonic excitations, which are combinations of electric and magnetic excitations, will be condensed. Most of the results are direct generalizations of the known gapping conditions of (2+1)d $\Z_N$ toric codes \cite{Juven, Levin, ChaoMing} with small modifications. 

For convenience, we repeat the 
boundary Lagrangian \eqref{eq:boundary_full},
\be
\L_0=-i\frac{N}{2\pi}  \left[ (\p_0 \p_x \vphi) \p_y \hat{\vphi}^{y}+(\p_0 \p_y \vphi) \p_x \hat{\vphi}^{x}\right].
\ee 
When $N$ is a prime number, the only possible topological boundaries are the smooth and rough ones as discussed in sections \ref{sec:mm_mm} and \ref{sec:ee_ee}. But when $N$ is composite, there is one more gapped boundary for each ordered decomposition $N=s\hat{s}$ with $s, \hat{s}\in\Z$. 
One can see this by further adding some auxiliary degrees of freedom decoupled from $\vphi$ and $\hat{\vphi}^i$:
\be
\L_{\text{aux}}^{[s\hat{s}]}=-i\frac{1}{2\pi}[(\p_0\p_x\chi)\p_y\hat{\chi}^y + (\p_0\p_y\vphi)\p_x \hat{\chi}^x],
\ee
which has the same form as $\L_0$ but with $N=1$. Such additional terms will not modify the bulk or the boundary physics, i.e., they will not affect the quasiparticle contents or the ground state degeneracy of the system. Then we further add the following gapping terms:
\be
\begin{split}
\L_g^{[s\hat{s}]}= & g_1\cos(N\hat{\vphi}^x-\hat{s}\hat{\chi}^x) +  g_2\cos(N\hat{\vphi}^y-\hat{s}\hat{\chi}^y) \\
& + g_3\cos (N\vphi+s\chi).
\end{split}
\label{eq:gap_chi}
\ee
One can easily check that at large $g_i$'s, the total boundary Lagrangian $\L_{\p \mathcal{M}}=\L_0+\L_{\text{aux}}+\L_g^{[s\hat{s}]}$ vanishes. The velocity terms of the $\vphi$, $\hat{\vphi}^i$ fields, and those of the $\chi$, $\hat{\chi}^i$ fields will also cancel out upon carefully chosing the velocity matrix for $\chi$ and $\hat{\chi}^i$. 

In general, such boundary conditions would lead to the combined condensation of quasiparticles of both electric and magnetic types, namely,  dyons, on the boundary. In particular, 
\be
(\hat{W}^x)^s\mid =
\exp \left[i \oint dx\ \p_x\hat{\chi}^x\right]=1
\ee
is condensed on the boundary. Similarly, $(\hat{W}^y)^{s}\,\,|$, $(W_{xz,x})^{\hat{s}}\mid$ and $(W_{yz,y})^{\hat{s}}\mid$ are also condensed. The contributions from these operators to the ground state degeneracy are summarized in table \ref{tab:topology}. The total ground state degeneracy is thus \be
\text{GSD}^{(s\hat{s})}=N^{l_x+l_y+2l_z}s^{-2}\hat{s}^{-1},
\ee
which recovers the results for the smooth and rough boundaries when $(s,\hat{s})=(N,1)$ and $(1,N),$ respectively.

\begin{table}[htbp]
    \centering
    \begin{tabular}{|c|c|c|c|}
    \hline
       \vtop{\hbox{\strut Non-commuting}\hbox{\strut \quad operators}}  &  Copies & \vtop{\hbox{\strut $\Z_{\#}$ Heisenberg}\hbox{\strut Algebras}}    & \vtop{\hbox{\strut Contribution}\hbox{\strut \quad to GSD}}\\ \hline
       $\hat{W}^x$, {$W_{xz,z}$}  &  $l_y$ &  $\Z_s$ & \multirow{2}{*}{$s^{l_y-1}N^{l_z}$} \\ \cline{1-3}
       $\hat{W}^x$, $W_{xy,y}$  &  $l_z$ & $\Z_N$  & \\ \hline
       $\hat{W}^y$, {$W_{yz,z}$}  &  $l_x$ & $\Z_s$ & \multirow{2}{*}{$s^{l_x-1}N^{l_z}$} \\ \cline{1-3}
       $\hat{W}^y$, $W_{xy,x}$  &  $l_z$ & $\Z_N$ & \\ \hline
       {$\hat{W}^z$}, $W_{yz,y}$  &  $l_x$ & $\Z_{\hat{s}}$ & \multirow{2}{*}{$\hat{s}^{l_x+l_y-1}$} \\ \cline{1-3}
       {$\hat{W}^z$}, $W_{xz,x}$  &  $l_y$ & $\Z_{\hat{s}}$ & \\ \hline
    \end{tabular}
    \caption{Summary of Wilson operators and their contributions to the ground state degeneracy when both boundaries are of type \eqref{eq:gap_chi}. In the first column of each row, the two Wilson operators are those that would have formed one copy of $\Z_N$ Heisenberg algebra if there were no boundaries. The second column describes the copy numbers of algebras formed by the operators in the first column, and the third column explains the properties of those algebras. The last column shows the contribution from those Wilson operators to the ground state degeneracy. 
    }
    \label{tab:topology}
\end{table}

Finally, we would like to comment that different dyonic quasiparticles can condense in different directions like what electric and magnetic planons did on the anisotropic boundary discussed in \ref{sec:em_em}, and different dyonic condensations can happen in different regions separated by domain walls, as discussed in \ref{subsec:geometry}. This combination of topological (dyon condensation) and geometric (domain wall arrangement) properties leads to the possibilities of a great number of different gapped boundary conditions in just this single X-cube model.

\section{Anomaly inflow}
\label{sec:anomaly}
 
In this section, we ask that starting from the boundary theory $\L_0$, whether the X-cube bulk theory is the unique bulk theory that cancels the anomaly. Spoiler: No. 

Recall that on the boundary we have
\be
\L_0=\frac{iK_{IJ}}{4\pi}\p_0\Phi_I\p_x\p_y \Phi_J,
\ee
where $K=-iN\sigma^y$. The symmetries were presented in \eqref{eq:enhanced}, and for convenience, we repeat the corresponding currents here, 
\be
J_{I,0}=-\frac{K_{IJ}}{2\pi}\p_x\p_y\Phi_J,\quad J_{I,xy}=0.
\ee
One can couple them to the background tensor gauge fields, which have the following $U(1)$ gauge transformations 
\be
(A_0^I,A_{xy}^I)\sim (A_0^I+\p_0\alpha_I,A_{xy}^I+\p_x\p_y \alpha_I).
\ee
The coupled Lagrangian is 
\be
\L_{2+1}[A]=\L_0-\frac{iK_{IJ}}{2\pi} A_0^I \p_x \p_y \Phi_J+\frac{iK_{IJ}}{4\pi} A_0^I A_{xy}^J,
\ee
where the last term is a local counterterm added to make the expressions look cleaner, but will not affect the analysis of anomaly.  It will not affect the discussions of anomalies. Under the $U(1)$ gauge transformations, $\L_{2+1}[A]$ becomes ($\Phi_I$ also needs to shift by $\alpha_I$, as dictated by \eqref{eq:A_phi}), 
\be
\L_{2+1}[A] \rightarrow  \L_{2+1}[A]+\frac{i K_{IJ}}{4\pi}[ A_{xy}^J \p_0\alpha_I -A_0^I \p_x\p_y \alpha_J].
\label{eq:anomaly}
\ee
In the equations above, we have used the fact that $K$ is antisymmetric: $K_{IJ}\p_0 \Phi_I \p_x\p_y\alpha_J=K_{IJ}\p_x\p_y\Phi_J \p_0\alpha_I,$ 
such that the terms linear in $\Phi$ cancel each other. We can easily observe that the anomaly in \eqref{eq:anomaly} cannot be removed by any local counterterm because of the relative sign in the middle. 
This anomaly can be canceled by coupling to a (3+1)d bulk, with the following gauge fields and gauge transformations 
\be
(A_0^I,A_{xy}^I,A_z^I)\sim (A_0^I+\p_0\alpha_I,A_{xy}^I+\p_x\p_y \alpha_I, A_z^I+\p_z\alpha_I). 
\label{eq:bulkgauge}
\ee
There are three field strengths for each flavor $I$,
\be
\begin{split}
 B^I&=\p_z A_{xy}^I-\p_x\p_y A_z^I,\\
 E^I_{xy}&=\p_0 A_{xy}^I-\p_x\p_y A_0^I,\\
 E_z^I&=\p_0 A_z^I-\p_z A_0^I. 
\end{split}
\ee
The bulk theory is then described by the following Lagrangian:
\be
\L_{3+1}=\frac{iK_{IJ}}{4\pi}(-A_0^I B^J +A_z^I E_{xy}^J +A_{xy}^I E_z^J).
\ee 
Under the gauge transformations \eqref{eq:bulkgauge}, the Lagrangian gets changed by
\be
\begin{split}
\L_{3+1} & \rightarrow \L_{3+1}  -\frac{iK_{IJ}}{4\pi}\p_z(A_{xy}^J \p_0\alpha_I -A_0^I \p_x\p_y\alpha_J )\\
\end{split}
\ee
which is simply a boundary term and cancels the anomaly we found in \eqref{eq:anomaly}.  One can easily see that $\L_{3+1}$ does not describe the X-cube phase, by looking at its symmetry operators. Consider $\L_{3+1}$ with periodic boundary conditions on all three directions, i.e., $T^3$. To distinguish from Wilson operators $W$ in the X-cube model, we denote symmetry operators here by $M$:
\be
\begin{split}
 M_{y}^I(x_1,x_2)&=\exp\left[i\int_{x_1}^{x_2}dx \oint dy  A^I_{xy}\right], \\
 M_{x}^I(y_1,y_2)&=\exp\left[i\int_{y_1}^{y_2}dy \oint dx  A^I_{xy}\right], \\
 M_{z}^I(x,y)&= \exp\left[i\oint dz  A^I_z\right]. \\
\end{split}
\ee
Note that the spatial dependence of $M_z^I(x,y)$ factorizes, $M_z^I(x,y)\equiv M_{z,x}^I(x)M^I_{z,y}(y)$. This is because when $B^I=0,$ we have
\be
\p_x\p_y\oint dz A_z^I =\oint dz\ \p_z A_{xy}^I=0.
\ee
So there are in general $(l_x+l_y-1)$ number of $M_z^I(x,y)$, with ``$-1$'' one from the factorization relation. On the other hand, $M_i^I$ with $i=x, y$  is independent of $z$ and there is one constraint $M_y^I(0,L_x)=M_x^I(0,L_y)$.  Therefore, for each flavor $I$, there are $(2l_x+2l_y-2)$ number of Wilson operators. 

The non-uniqueness of the anomaly-cancelling bulk theories is common in systems with subsystem symmetries because there are various ways to extend the boundary foliation into the bulk \cite{anomaly,private}.


\section{Summary and discussion}
\label{sec:discussion}

In this work, the boundary theory of X-cube model was studied from a continuum field theory perspective. This boundary theory is a generalization of the boundary theory of (2+1)d $\Z_N$ toric code to one higher dimension. We examined its symmetries, algebras, and extracted bulk quasiparticle statistics from the boundary fields. We analyzed general possible gapped boundary conditions that either preserve or break the rotation/translation symmetries on the boundary, and their corresponding ground state degeneracies. In particular, both the extensive and the constant parts in the ground state degeneracy can vary with different choices of boundary conditions. We further discussed the anomaly inflow from a single boundary and found the bulk theory is not unique. 

We would like to comment that we have described the gapped boundary conditions in terms of cosine potentials for the boundary theory. Alternatively, one might consider directly using Dirichlet boundary
conditions of the tensor gauge fields. For the simplest smooth and rough boundary cases, this is possible. For example, the smooth boundary can be easily achieved by choosing
$A_0 = 0, A_{xy} = 0$, while the rough boundary can be obtained by imposing $\hat{A}^{z(xy)}_0 = 0,$ and $\hat{A}^{xz} = 0 = \hat{A}^{yz}.$ However, for general gapped boundaries breaking rotation/translation
symmetries or those with dyon condensations, it is not obvious how one can formulate them using Dirchlet boundary conditions. For instance, the anisotropic boundary in the continuum requires $\oint dx A_{xy}$ on the boundary to be
trivial, while $\oint dy A_{xy}$ remains nontrivial. Consequently, one cannot simply take $A_{xy}$ = 0 on the
boundary. 
As for boundaries with general dyon condensations, the Wilson operators need satisfy equations of the format
$\langle W^s \rangle =$const., with $s> 1$. From this perspective, our formalism with boundary dynamical fields $\vphi$ and $\hat{\vphi}^i$ are more natural.

This work leads naturally to the following questions that we will leave for future work.

(i) We have focused on the $z=$ constant type of boundaries. Generally there are other types of planar boundaries such as ones with $x+y=$ constant. However, such boundary theories keep less symmetries and cannot be analyzed by a straightforward extension of this work. For example consider the $x+y=0$ boundary. Following the similar procedures that lead to \eqref{eq:boundary}, we find the variation of action under a gauge transformation vanishes when the boundary condition \eqref{eq:temporal} is imposed. Extending this as a temporal gauge into the bulk, we 
again solve for the relations between the gauge fields and $(\vphi, \hat{\vphi}^i)$ as in \eqref{eq:A_phi}, and the Lagrangian reduces to 
\be
\L=\frac{iN}{2\pi}\frac{1}{2} \p_u\left[ (\p_u\vphi) \p_0\p_z\hat{\vphi}^z +(\p_v\vphi)  \p_0\p_z (\hat{\vphi}^y-\hat{\vphi}^x)\right].
\ee
The second term in the square bracket is a pure boundary term, but the first term contains double derivatives $\p_u^2$. This complication reflects the fact that the chosen boundary is not compatible with the cubic symmetry. To correctly describe the $(110)$-type boundary, one needs to examine a staircase-shaped termination of the bulk theory, which will be left for future work.

(ii) We have ignored the perturbative relevance/irrelevance of the gapping terms, as we are only interested in whether local interactions can gap out the boundary when they are strong enough, analogous to the case of topological orders \cite{Juven,Levin,ChaoMing}. It would also be interesting  to thoroughly examine the RG flows of such terms. The concept of renormalization in fractonic systems is a subtle one as discussed in \cite{PhysRevB.89.075119,PhysRevB.93.045127,10.21468/SciPostPhys.6.4.041,PhysRevX.8.031051,10.21468/SciPostPhys.6.1.015,SHIRLEY2019167922,PhysRevResearch.2.033021}, and is beyond the scope of the current paper. However, we expect that the procedure mapped out in Ref.  \cite{Ethan} could be utilized to examine our system of interest and the gapping terms discussed in the previous sections will all be perturbatively relevant.

(iii) We have written the boundary theory in terms of a  $K$-matrix. One might wonder if more exotic fracton phases of matter can be generated in this way, by allowing $K$ to be a general anti-symmetric integer matrix, and extending back into the bulk. Unfortunately for a general $d\times d$-dimensional anti-symmetric matrix $K$ of rank $2n$, one can always linearly transform it to the factorized form of $\oplus_{i=1}^n N_i i\sigma^y\oplus\mathbb{0}_{d-2n}$, where $\mathbb{0}_{d-2n}$ is a $(d-2n)\times (d-2n)$-dimensional block of zeros. The details should be examined more carefully, but we do not expect super exciting consequences with a general $K$. 

(iv) One can further study the relationship between the boundary and the bulk theories by examining the relations between the boundary energy spectrum and the entanglement spectrum \cite{PhysRevLett.101.010504} in the bulk. The entanglement entropy for fracton phases have been investigated in references \cite{Entanglement,PhysRevB.97.144106,Shirley_2019}.

\section*{Acknowledgement}
We are grateful to Ho Tat Lam for carefully reading the draft and making important comments. We thank Daniel Bulmash for very helpful suggestions on the draft and Shu-Heng Shao for inspiring discussions. Zhu-Xi thanks Po-Shen Hsin for sharing his unpublished work on foliated field theory; Arpit Dua, Tarun Grover, Joseph Sullivan and Cenke Xu for conversations on coupled wire constructions; and Xie Chen, Jeongwan Haah, Wenjie Ji, Ethan Lake, Shang Liu, Kevin Slagle, Xiao-Chuan Wu, Ruben Verresen, Sagar Vijay, Ashvin Vishwanath and Xiao-Gang Wen for general discussions. We thank the first referee of PRB for pointing out an error in appendix A. 
ZXL is supported by the Simons Collaborations on Ultra-Quantum Matter, grant 651457 from the Simons Foundation. The work of AK was supported, in part, by the U.S.~Department of Energy under Grant DE-SC0022021 and by a grant from the Simons Foundation (Grant 651678, AK). The work of HYS was supported by a grant from the Simons Foundation (Grant 651678, AK).

\appendix

\section{Symmetry groups and their irreducible representations}
\label{app:repn}

In this appendix, we present the irreducible representations of the symmetry groups of the system and relate them to the bulk and the boundary fields. The bulk result is a review of that in \cite{SS3}.

Since the bulk of the system lives on a cube lattice, all the fields can be labeled by irreducible representations of the cubic group. In particular, the orientation-preserving subgroup of the cubic group is $S_4$, on which we will focus. Following the notations in \cite{SS3}, we label irreducible representations of $S_4$ by their dimensions: the trivial representation $\textbf{1}$, the sign representation $\textbf{1}'$, the two-dimensional representation $\textbf{2}$, the standard representation $\textbf{3}$, and another three dimensional representation $\textbf{3}' = \textbf{1}'\otimes \textbf{3}$. 

It is useful to decompose the irreducible representations of $SO(3)$ furnished by symmetric traceless tensors into the irreducible representations of $S_4$.  The first few are
\begin{equation}
\begin{split}
    SO(3) & \supset S_4 \\
    \textbf{1} & = \textbf{1} \\
    \textbf{3} & = \textbf{3} \\
    \textbf{5} & = \textbf{2}\oplus \textbf{3}' \\
    \textbf{7} & = \textbf{1}'\oplus \textbf{3} \oplus \textbf{3}'\\
    \textbf{9} & = \textbf{1}\oplus \textbf{2} \oplus \textbf{3} \oplus \textbf{3}'.\\
\end{split}
\label{eq:S4}
\end{equation}
We will follow the standard conventions for indices: $(ab)$ symmeterizes indices and $[ab]$ antisymmeterizes indices. Using the decompositions above, the representations of $S_4$ can be expressed in terms of the following tensors:
\begin{equation}
\begin{split}
& \textbf{1} : S \\
    &\textbf{1}' : T_{(ijk)} ;\quad  i\neq j \neq k \\
    &\textbf{2}: B_{[ij]k} ;\quad  i \neq j \neq k ; \\
    & \quad \quad \quad \quad \quad B_{[ij]k} + B_{[jk]i} + B_{[ki]j} = 0\\
    &\textbf{2}: B_{i(jk)};\quad i \neq j \neq k; \\
    & \quad \quad \quad \quad \quad B_{i(jk)} + B_{j(ki)} + B_{k(ij)} = 0\\
    &\textbf{3}: V_i \\
    &\textbf{3}': E_{ij};\quad  i\neq j; \quad E_{ij} = E_{ji} \\
\end{split}
\end{equation}
The two different expressions for $\textbf{2}$ come from the $\textbf{2}$ in $\textbf{3}\otimes \textbf{3} = \textbf{1}\oplus \textbf{2}\oplus \textbf{3}\oplus \textbf{3}$ and  $\textbf{3}\otimes \textbf{3}' = \textbf{1}'\oplus \textbf{2}\oplus \textbf{3}\oplus \textbf{3}'$, respectively, and can be related to each other.

On the boundary, we have a square lattice with the orientation-preserving subgroup being $\mathbb{Z}_4$, corresponding to the fourfold rotation with respect to an axis perpendicular to the square lattice and passing through a site. $\mathbb{Z}_4$ has four one-dimensional irreducible representations labeled by $\bm{1}_q$ with $q=0, 1, 2, 3.$ 

Again we decompose the irreducible representations of SO(3) in terms of these irreducible representations of $\Z_4$. 
The first step is $SO(3)\rightarrow SO(2)$. There is one irreducible representation  $\bm{2q+1}$ of $SO(3)$ for each integer $q$, which can be decomposed into representations of $SO(2)$:
\begin{equation}
\begin{split}
SO(3) & \supset SO(2)\\
\bm{2q+1} & = \bm{2}_q \oplus \bm{2}_{q-1} \oplus \cdots \oplus \bm{2}_1 \oplus \bm{1}_0,
\end{split}
\end{equation}
where $\bm{2}_q$ corresponds to the two-dimensional rotation matrix with rotation angle $q\theta$, while $\bm{1}_0$ is an one-dimensional matrix, i.e., a number, $1$.

In the second step, we combine the equation above with the decomposition in \eqref{eq:S4}, we arrive at
\begin{equation}
\begin{split}
S_4 & \rightarrow SO(2)\\
\bm{1} & = \bm{1}_0\\
\bm{3} & = \bm{2}_{1} \oplus \bm{1}_0\\
\bm{2} \oplus \bm{3}' & = \bm{2}_{2} \oplus \bm{2}_{1} \oplus \bm{1}_0. \\
\bm{1}'\oplus \bm{3} \oplus \bm{3}' 
& =\bm{2}_3 \oplus \bm{2}_2 \oplus \bm{2}_1\oplus \bm{1}_0.\\ 
\end{split}
\label{eq:S4_SO2}
\end{equation}
The second and the last equations in \eqref{eq:S4_SO2} can be combined to give
\be
\bm{1}' \oplus \bm{3}'  =\bm{2}_3 \oplus \bm{2}_2.
\label{eq:temp1}
\ee

We further decompose
\begin{equation}
\begin{split}
SO(2)  & \supset  U(1)\\
\bm{2}_q & = \bm{1}_q \oplus \bm{1}_{-q}, 
\end{split}
\end{equation}
obtained from the diagnoalization of the two-dimensional rotation matrix. Going from $U(1)$ to $\mathbb{Z}_4$, we simply take $\bm{1}_q=\bm{1}_{q \mod 4}$. Therefore, equations \eqref{eq:S4_SO2} and \eqref{eq:temp1} result in
\begin{equation}
\begin{split}
S_4 & \supset \mathbb{Z}_4\\
\bm{1} & = \bm{1}_0\\
\bm{3} & = \bm{1}_{1} \oplus \bm{1}_{3} \oplus \bm{1}_0\\
\bm{2}\oplus \bm{3'} & = \bm{1}_{2} \oplus \bm{1}_{2} \oplus \bm{1}_{1} \oplus \bm{1}_{3} \oplus \bm{1}_0.\\
\bm{1}'\oplus \bm{3}' & =\bm{1}_3 \oplus \bm{1}_1 \oplus \bm{1}_2 \oplus \bm{1}_2.\\
\end{split}
\label{eq:S4_Z4}
\end{equation}
Next, we multiply both sides of the second equation in \eqref{eq:S4_Z4} by $\bm{1}'$ and make use of the fusion rule 
$\bm{3}'=\bm{3}\otimes \bm{1}'$ in $S_4$. Assuming the $\bm{1}'$ representation in $S_4$ is identified with the $\bm{1}_a$ representation in $\Z_4$, we have
\be
\bm{3}'=\bm{1}_{a+1}\oplus \bm{1}_{a+3} \oplus \bm{1}_a.
\ee
Plug it back into the last line of \eqref{eq:S4_Z4}, we arrive at the following consistency equation for $a$:
\be
\bm{1}_a\oplus \bm{1}_{a+1}\oplus \bm{1}_{a+3} \oplus \bm{1}_a = \bm{1}_3 \oplus \bm{1}_1 \oplus \bm{1}_2 \oplus \bm{1}_2.
\ee
It is then straightforward to see that $a=2$, and $\bm{3}'=\bm{1}_3\oplus \bm{1}_1\oplus \bm{1}_2.$ Finally using the third equation in \eqref{eq:S4_Z4}, we arrive at
\be
\bm{2}=\bm{1}_2\oplus \bm{1}_0.
\ee
Since the boundary Lagrangian $\L_0\propto (\p_0\vphi)(\p_x\p_y \hat{\vphi}^{z(xy)})$ is invariant under the $\Z_4$ and $\p_x\p_y$ transforms under $\bm{1}_2$, $\hat{\vphi}^{z(xy)}$ must transform under $\bm{1}_2$. The remaining degree of freedom,  $\hat{\varphi}^{x(yz)}-\hat{\varphi}^{y(xz)}$ thus transforms under the  $\bm{1}_0$ representation of $\Z_4$.

\section{Coupling to background fields}
\label{sec:background}

In this appendix, we further explain the interpretation of $\rho_{I}$ as the dipole density by coupling to background tensor fields. The background gauge fields for the $\Z_N$ subsystem symmetries of the X-cube model and their couplings to the bulk dynamical fields were discussed in \cite{anomaly}. The coupling is 
\be
\begin{split}
\L_J=\frac{i}{4\pi}[& -A_{ij}(\p_0\hat{a}^{ij}-\p_k\hat{a}_0^{k(ij)})-A_0(\p_i\p_j\hat{a}^{ij})\\
& +\hat{A}^{ij}(\p_0 a_{ij}-\p_i\p_ja_0)+\hat{A}_0^{k(ij)}(\p_i a_{jk}-\p_j a_{ik})]
\end{split}
\ee
where $(a_0, a_{ij})$ and $(\hat{a}_0^{k(ij)},\hat{a}^{ij})$ are the $U(1)$ background gauge fields. They can be Higgsed down to $\Z_N$ by coupling to dynamical tensor or scalar fields. In the temporal gauge of the dynamic gauge fields, the above coupling reduces to 
\be
\L_J=\frac{i}{4\pi}[\p_k\hat{\vphi}^{k}(\p_0 a_{ij}-\p_i\p_ja_0)-\p_i\p_j\vphi(\p_0\hat{a}^{ij}-\p_k\hat{a}_0^{k(ij)})],
\ee
where we have also used \eqref{eq:A_phi}. Foucsing on the terms containing the temporal components of the background field, we reduce them to the boundary term
\be
\frac{i}{2\pi}\p_z[\hat{a}_0^{z(xy)}(\p_x\p_y\vphi)-a_0(\p_x\p_y\hat{\vphi}^z)].
\label{eq:density}
\ee
We thus notice that
$\p_x\p_y \hat{\vphi}^z$ describes the density of fractons on the boundary, while $\p_x\p_y\vphi$ describes the density of $z$-lineons on the boundary.

\bibliography{ref.bib}

\appendix
\begin{widetext}
\end{widetext}
\end{document}